\documentstyle[a4p,12pt,oldlfont,twoside,cite,epsfig,rotating]{article}
\newcommand{\aem    }{\mbox{$\alpha$ }}
\newcommand{\aemsq  }{\mbox{$\aem^2$}}

\newcommand{\Z      }[4]{\mbox{$#1\,\pm #2\,^{+\,#3}_{-\,#4} $}}
\newcommand{\qsq    }{\mbox{$Q^{2}$}}
\newcommand{\psq    }{\mbox{$P^{2}$}}

\newcommand{\ft     }{\mbox{$F_{2}^{\gamma}$}}
\newcommand{\ftmi   }{\mbox{$F_{2,i}^{\gamma}$}}
\newcommand{\ftmc   }{\mbox{$F_{2,\rm MC}^{\gamma}$}}
\newcommand{\ftn    }{\mbox{$F_{2}^{\gamma}/\aem$}}
\newcommand{\fl     }{\mbox{$F_{\rm L}^{\gamma}$}}
\newcommand{\ftxq   }{\mbox{$\ft(x,\qsq)$}}

\newcommand{\ftxqp  }{\mbox{$\ft(x,\qsq,\psq\ne 0)$}}
\newcommand{\flxq   }{\mbox{$\fl(x,\qsq)$}}
\newcommand{\grg    }{\mbox{$\gamma\gamma$}}
\newcommand{\gsg    }{\mbox{$\gamma^{\star}\gamma$}}
\newcommand{\gsgs   }{\mbox{$\gamma^{\star}\gamma^{\star}$}}
\newcommand{\avqsq  }{\mbox{$\langle\qsq\rangle$}}
\newcommand{\pdf    }{\mbox{${\rm f}_{\rm q,\gamma}(x,\qsq)$}}

\newcommand{\gev    }{\mbox{$\rm GeV$}}
\newcommand{\gevsq  }{\mbox{$\rm GeV^2$}}
\newcommand{\zn     }{\mbox{$\rm Z^0$}}

\newcommand{\epem   }{\mbox{$\rm e^+e^-$}}
\newcommand{\ww     }{\mbox{$\rm W^+W^-$}}
\newcommand{\qq     }{\mbox{$\rm q\overline{q}$}}
\newcommand{\tptm   }{\mbox{$\rm \tau^+\tau^-$}}

\newcommand{\znhad  }{\mbox{$\zn\rightarrow$ hadrons}}
\newcommand{\znhads }{\mbox{$\zn\rightarrow$ had.}}
\newcommand{\wwhad  }{\mbox{$\ww\rightarrow$ hadrons}}
\newcommand{\gsgshad}{\mbox{$\gsgs\rightarrow$ hadrons}}
\newcommand{\gsgshads}{\mbox{$\gsgs\rightarrow$ had.}}
\newcommand{\zntau  }{\mbox{$\zn\rightarrow \tau^+\tau^-$}}
\newcommand{\ggtau  }{\mbox{$\gamma^{\star}\gamma\rightarrow\tau^+\tau^-$}}
\newcommand{\ggmu   }{\mbox{$\gamma^{\star}\gamma\rightarrow\mu^+\mu^-$}}
\newcommand{\ggep   }{\mbox{$\gamma^{\star}\gamma\rightarrow\epem$}}
\newcommand{\th     }{\mbox{$\theta$}}
\newcommand{\ph     }{\mbox{$\phi$}}

\newcommand{\etag   }{\mbox{$E_{\rm tag}$}}
\newcommand{\ea     }{\mbox{$E_{\rm a}$}}

\newcommand{\nch    }{\mbox{$N_{\rm ch}$}}
\newcommand{\eb     }{\mbox{$E_{\rm b}$}}
\newcommand{\ttag   }{\mbox{$\theta_{\rm tag}$}}
\newcommand{\qzm    }{\mbox{$\langle \qsq \rangle$}}
\newcommand{\pzm    }{\mbox{$\langle \psq \rangle$}}

\newcommand{\etout  }{\mbox{$E_{\rm T}^{\rm out}$}}
\newcommand{\wfd    }{\mbox{$W_{\rm vis}$}}
\newcommand{\wrec   }{\mbox{$W_{\rm rec}$}}
\newcommand{\wcor   }{\mbox{$W_{\rm cor}$}}
\newcommand{\wfdsq  }{\mbox{$W_{\rm vis}^2$}}
\newcommand{\wrecsq }{\mbox{$W_{\rm rec}^2$}}
\newcommand{\wcorsq }{\mbox{$W_{\rm cor}^2$}}

\newcommand{\xfd    }{\mbox{$x_{\rm vis}$}}
\newcommand{\xrec   }{\mbox{$x_{\rm rec}$}}
\newcommand{\xcor   }{\mbox{$x_{\rm cor}$}}
\newcommand{\edist  }{\mbox{$E_{\rm for}/E_{\rm tot}$}}
\newcommand{\etdist }{\mbox{$E_{\rm T}^{\rm out}/E_{\rm tot}$}}

\newcommand{\pt     }{\mbox{$p_{\rm t}$}}
\newcommand{\act    }{\mbox{$|\cos\theta|$}}
\newcommand{\restr  }{\mbox{$\sigma_{p_{\rm t}}/p_{\rm t}=\sqrt{0.02^2+(0.0015\,\pt)^2}$}}
\newcommand{\kt     }{\mbox{$k_{\mathrm{t}}(\rm dyn)$}}
\newcommand{\chmodels}{\mbox{${\chi^2}$}}
\def\etal{{\it et al.}}

\begin{document}

\begin{titlepage}

\bigskip\def\thefootnote{}
\center{\large EUROPEAN ORGANISATION FOR NUCLEAR RESEARCH}
\bigskip
\flushright{CERN-EP-2000-082\\
            08 June 2000}
\bigskip\bigskip
\begin{center}
  {\huge\bf\boldmath
    \center{ Measurement of the}\vspace{-0.75cm}
    \center{ Low-$x$ Behaviour of the}\vspace{-0.5cm}
    \center{Photon Structure Function \boldmath\ft}
  }
\end{center}

\bigskip\bigskip 
\begin{center}{{\bf \LARGE The OPAL Collaboration}}\end{center}
\bigskip\bigskip
\begin{abstract}
  The photon structure function \ftxq\ has been measured using data
  taken by the OPAL detector at \epem\ centre-of-mass energies of
  91~\gev, 183~\gev\ and 189~\gev, in $Q^2$ ranges of 
  1.5--30.0 GeV$^2$ (LEP1), 
  and 7.0--30.0 GeV$^2$ (LEP2), 
  probing lower values of $x$ than ever before.
  Since previous OPAL analyses,
  new Monte Carlo models and new methods,
  such as multi-variable unfolding, have been introduced, 
  reducing significantly the model dependent systematic errors in 
  the measurement.
\end{abstract}  
\bigskip\bigskip\bigskip
\begin{center}{\large
(Submitted to European Physical Journal C)
}\end{center}
\end{titlepage}
%
%

\begin{center}{\Large        The OPAL Collaboration
}\end{center}\bigskip
\begin{center}{
G.\thinspace Abbiendi$^{  2}$,
K.\thinspace Ackerstaff$^{  8}$,
C.\thinspace Ainsley$^{  5}$,
P.F.\thinspace Akesson$^{  3}$,
G.\thinspace Alexander$^{ 22}$,
J.\thinspace Allison$^{ 16}$,
K.J.\thinspace Anderson$^{  9}$,
S.\thinspace Arcelli$^{ 17}$,
S.\thinspace Asai$^{ 23}$,
S.F.\thinspace Ashby$^{  1}$
% ...
}\end{center}\bigskip
\bigskip
$^{  1}$School of Physics and Astronomy, University of Birmingham,
Birmingham B15 2TT, UK
\newline
$^{  2}$Dipartimento di Fisica dell' Universit\`a di Bologna and INFN,
I-40126 Bologna, Italy
\newline
$^{  3}$Physikalisches Institut, Universit\"at Bonn,
D-53115 Bonn, Germany
\newline
% ...
$^{  5}$Cavendish Laboratory, Cambridge CB3 0HE, UK
\newline
% ...
$^{  8}$CERN, European Organisation for Nuclear Research,
CH-1211 Geneva 23, Switzerland
\newline
$^{  9}$Enrico Fermi Institute and Department of Physics,
University of Chicago, Chicago IL 60637, USA
\newline
% ...
$^{ 16}$Department of Physics, Schuster Laboratory, The University,
Manchester M13 9PL, UK
\newline
$^{ 17}$Department of Physics, University of Maryland,
College Park, MD 20742, USA
\newline
% ...
$^{ 22}$Department of Physics and Astronomy, Tel Aviv University,
Tel Aviv 69978, Israel
\newline
$^{ 23}$International Centre for Elementary Particle Physics and
Department of Physics, University of Tokyo, Tokyo 113-0033, and
Kobe University, Kobe 657-8501, Japan
\newline
% ...
Budapest, Hungary.
%
%
\section{Introduction}
\begin{figure}[htb]
\begin{center}
\mbox{\epsfig{bbllx=0,bblly=200,bburx=595,bbury=580,file=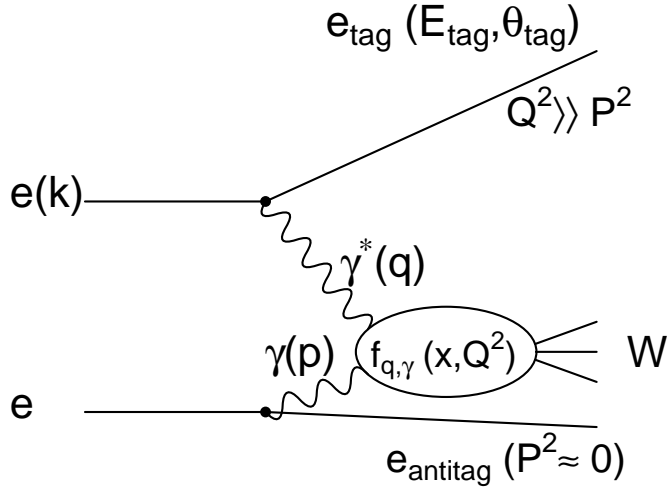,height=6.5cm}}
\caption{\label{fig:dis} 
 Deep inelastic electron-photon scattering.}
\end{center}
\end{figure}

The measurement of the hadronic photon structure function
\ftxq\ is a classic test of QCD predictions~\cite{ref:QPM,ref:F2QCD,ref:Berger}.
The structure function of the photon differs from 
that of the proton because the photon can couple directly 
to quark charges, as well as fluctuate into a hadronic state.
The value of $F_{2}^{\mbox{\tiny proton}}(x,Q^2)$ exhibits a clear rise 
towards low values of Bjorken $x$~\cite{ref:ZEUS,ref:H1}.
Some theoretical models predict a similar rise in the photon 
structure function~\cite{ref:GRV,ref:SaS1D,ref:WHIT}, 
while other models do not
require such a rise~\cite{ref:fkp}. Experimentally, 
a rise at low-$x$ in the photon 
structure function has neither been observed nor 
excluded~\cite{ref:PLUTO,ref:JADE,ref:TASSO,ref:TPC,ref:AMY,ref:TOPAZ,
ref:OPAL,ref:DELPHI,ref:L3,ref:ALEPH}.
The photon structure function is studied at LEP using samples of
events of the type 
$\epem \rightarrow \epem\gsg \rightarrow \epem +\mbox{hadrons}$. 
The analysis presented here uses
single-tagged events (from here on referred to as \gsg\ events), 
which means that only one of the
scattered beam electrons\footnote{For conciseness, positrons are also referred to as electrons.}
is observed in the detector. 
These events can be regarded as
deep inelastic scattering of an electron off a quasi-real target photon, 
and the flux of quasi-real photons can be calculated using the
equivalent photon approximation~\cite{ref:QPM}.
Figure~\ref{fig:dis} shows a diagram of this reaction, in which
$k$ is the four-vector of the incoming electron which radiates the 
virtual photon, and $q$ and $p$ are the four-vectors of the virtual 
photon and the quasi-real photon, respectively.
The symbol \pdf\ represents the parton densities of the 
quasi-real photon.

The cross-section for deep inelastic electron-photon scattering can be 
written in terms of structure functions as~\cite{ref:Berger}
 \begin{equation}
  \frac{\mbox{d}^2\sigma_{\rm e\gamma\rightarrow e X}}{\mbox{d}x\mbox{d}Q^2}
   =\frac{2\pi\aemsq}{x\,Q^{4}}
  \left[\left( 1+(1-y)^2\right) \ftxq - y^{2} \flxq\right]
 \label{eqn:Xsect}
 \end{equation}
where $\qsq=-q^2$ is the negative value of the four-momentum squared 
of the virtual probe photon, \aem\ is the fine structure 
constant and $x$ and $y$ are the usual dimensionless 
deep inelastic scattering variables, defined by
  \begin{equation}
    x=\frac{\qsq}{2p \cdot q},\ y=\frac{p \cdot q}{p \cdot k}. 
  \end{equation}
In the kinematic region studied in this paper, $y^2\ll 1$, 
so the contribution from the longitudinal photon structure function \fl\
in Equation~\ref{eqn:Xsect} is neglected.

In contrast to measurements of the proton structure function, here
the energy of the target particle is not known. In consequence,
the kinematics cannot be fully determined without measuring the
hadronic final state, which is only partially observed in the
detector. This leads to a dependence of the \ft\ measurement on 
Monte Carlo modelling of the hadronic final state, which enters
when an unfolding process is used to relate the visible
distributions to the underlying $x$ distribution.

The analysis presented here uses OPAL data collected during the years 1993--1995,
1997 and 1998, at \epem\ centre-of-mass energies of 91~\gev\ (LEP1),
and 183~\gev\ and 189~\gev\ (LEP2). This is the first OPAL measurement of 
the four-flavour photon structure function 
using the 183~\gev\ and 189~\gev\ data,
though the charm contribution to \ft\ has already been 
measured~\cite{ref:charm}.
In this analysis new Monte Carlo programs and improved unfolding 
methods have been introduced, which are also used in the 
re-analysis of the LEP1 data.
%
%
\section{The OPAL detector}
 The OPAL detector is described in detail elsewhere~\cite{ref:opalnim};
 only the subdetectors which are most relevant for this analysis, namely 
 the low angle electromagnetic calorimeters and the tracking devices, are 
 detailed below.
 The OPAL detector has a uniform magnetic field of 0.435~T 
 along the beam direction throughout the central tracking
 region, with electromagnetic and hadronic calorimetry and muon chambers
 outside the coil.\par
 The forward detectors (FD) cover the $\theta$ region from 60 to
 140~mrad at each
 end of the OPAL detector\footnote{In the OPAL right-handed coordinate system the 
 $x$-axis points towards the
 centre of the LEP ring, the $y$-axis points upwards and the $z$-axis points in
 the direction of the electron beam.  The polar angle \th\ and the 
 azimuthal angle \ph\ are defined with respect to the $z$-axis
 and $x$-axis, respectively.}. 
 They consist of cylindrical lead-scintillator calorimeters    
 with a depth of 24 radiation lengths ($X_0$) divided azimuthally 
 into 16 segments. The energy resolution for electromagnetic showers 
 is $18\%/\sqrt{E}$, where $E$ is in \gev.  
 An array of three planes of proportional tubes buried in the 
 calorimeter at a depth of 4~$X_0$ provides a precise shower position 
 measurement, with a typical resolution of 3--4~mm, corresponding to 
 2.5~mrad in \th, and less than 3.5~mrad in \ph.\par
%
 The small-angle silicon tungsten luminometer (SW)
 covered the region in \th\ from 25 to
 60~mrad from 1993--1995. 
 For LEP2 running, a radiation shield was installed
 which moved the lower edge of the useful SW acceptance to 33~mrad.
 The SW detector contains 19 layers of silicon alternating 
 with tungsten. Each of the 16 azimuthal wedges is 
 divided into 64 pads for positional measurement. 
 The energy resolution of the SW detector is about $24\%/\sqrt{E}$ 
 at LEP1 energies, and about $6\%$ at LEP2.\par  
 Charged particles are detected by a silicon microvertex detector, a drift
 chamber vertex detector, and a jet chamber.
 Outside the jet chamber, but still inside the magnetic field, lies
 a layer of drift chambers whose purpose is to improve the track
 reconstruction in the $z$-coordinate.
 The resolution of the transverse\footnote{Transverse is always defined with 
 respect to the $z$-axis.}
 momentum for charged particles is \restr\
 for $\act < 0.7$ , where $p_{\rm t}$ is in GeV, 
 and degrades for higher values of \act.
 Tracks are accepted up to a limit of $\act< 0.9622$.\par
 Both ends of the OPAL detector are equipped with
 electromagnetic endcap calorimeters covering the 
 range from 200 to 630~mrad in polar angle. They are homogeneous
 devices composed of arrays of lead-glass blocks of
 $9.2 \times 9.2$~cm$^2$ cross-section and typically 22 $X_0$
 in depth, giving good shower containment.
 In the central region, outside the solenoid, is the electromagnetic 
 barrel calorimeter of similar construction.
 \par
 The deep inelastic scattering events are triggered with high efficiency 
 by the large energy deposits of the scattered electron in the 
 forward calorimeters (FD and SW) 
 and by charged particles seen in the tracking devices.
%
\section{Kinematics and data selection}
To measure \ftxq, the distribution of events in $x$ and \qsq\ is needed. These 
variables,
illustrated in Figure~\ref{fig:dis},
are related to the experimentally measurable quantities
\begin{equation}
  \qsq = 2\,\eb\,\etag\,(1- \cos\ttag)
  \label{eqn:qsq}
\end{equation}
and
\begin{equation}
  x = \frac{\qsq}{\qsq+W^2+\psq}
  \label{eqn:Xcalc}
\end{equation}
where \eb\ is the energy of the beam electrons, \etag\ and \ttag\ are the 
energy and polar angle of the deeply inelastically scattered 
(or `tagged') electron, $W^2$ is the invariant mass
squared of the hadronic final state and $\psq=-p^2$ is the negative 
value of the virtuality squared of the quasi-real photon.
The requirement that the associated electron is not 
seen in the detector ensures that $\psq \ll \qsq$, so \psq\ is
neglected when calculating $x$ from Equation~\ref{eqn:Xcalc}.
The electron mass is neglected throughout. 

Three samples of events are studied in this analysis, 
classified according to the centre-of-mass energy
and the subdetector in which the tagged electron is observed. 
Data from LEP1 are used, with \epem\ centre-of-mass energies
close to the \zn\ mass, as well as 
LEP2 data with centre-of-mass energies of 183~\gev\ and 189~\gev.
Electrons are tagged using the SW detector at all centre-of-mass 
energies, since
accessing the lowest possible $x$ region requires measuring the 
electrons scattered at the lowest possible angles.
The FD is only used for the LEP1 data, to provide a sample 
in the same range of \qsq\ as the LEP2 sample, but using a 
different subdetector for tagging the scattered electron.
The three samples used are termed LEP1 SW, LEP1 FD and LEP2 SW.
Each sample is further split into two bins of \qsq.

Events are selected by applying cuts on the scattered electrons and on 
the hadronic final state. The cuts are listed in Table~\ref{tab:cuts}.
A tagged electron within the clear acceptance of SW or FD is selected by 
requiring $\etag \ge 0.75\,\eb$ at LEP1 or $\etag \ge 0.775\,\eb$ at LEP2.
This cut effectively eliminates events originating from random 
coincidences between
off-momentum\footnote{Off-momentum electrons originate from beam gas
interactions  far from the OPAL interaction region and are 
deflected into the detector by the focusing
quadrupoles. They appear predominantly in the plane of the accelerator.} 
beam electrons faking a tagged electron and untagged \grg\ events.
The cut is higher at LEP2 because of the larger off-momentum 
background in the 183~\gev\ data. 
To ensure that the virtuality of the target photon is small, 
the highest energy cluster in the hemisphere opposite the tagged 
electron must have an energy $\ea \le 0.25\,\eb$ (the anti-tag condition).
To reject background from \gsg\ events with leptonic final states, 
the number of tracks in the event 
passing quality cuts and originating from the hadronic final state, 
\nch, must be at least three, of which at least two tracks must not be 
identified as electrons. 
 The tracks and the calorimeter clusters are reconstructed using 
 standard OPAL techniques~\cite{ref:tracks} which avoid double counting
of particles which produce both tracks and clusters.
Finally, the visible invariant mass \wfd\ of the hadronic
system based on tracks and calorimeter clusters and
including the contribution from energy measured in the forward
calorimeters (FD and SW) is required to be in the range 
$2.5\ \gev \le \wfd \le 40~\gev$ for the LEP1 sample and 
$2.5\ \gev \le \wfd \le 60~\gev$ for the LEP2 sample. 
The distribution of \wfd\
extends to higher values at LEP2 because more energy is available 
from the beam electrons.
In addition, the background from \znhad, which the maximum \wfd\ cut is 
designed to reject, is lower at LEP2 than at LEP1.

The trigger efficiencies were evaluated from the data using sets of 
separate triggers and found to be larger than $99\%$ for all of the samples. 

The cuts applied to each sample
are listed in Table~\ref{tab:cuts} and shown as dotted lines in
Figures~\ref{91swgen1}--\ref{189swgen2}.
The numbers of events in each sample passing the
cuts, the integrated luminosities and
the \qsq\ ranges 
are listed in Table~\ref{tab:samples}.
The luminosity for the LEP1 data is lower for the SW sample than for the 
FD sample because the SW sample does not include data from 1995.
\begin{table}[htb]
  \begin{center}\begin{tabular}{|c|c|c|c|}\hline
    cut               & LEP1 SW & LEP1 FD         & LEP2 SW \\\hline
    \etag/\eb\ min  & \multicolumn{2}{c|}{0.75} & 0.775   \\\hline
    \ttag\ min [mrad] & 27      & 60              & 33.25   \\\hline
    \ttag\ max [mrad] & 55      & 120             & 55      \\\hline
    \ea/\eb\ max      & \multicolumn{3}{c|}{0.25}           \\\hline
    \nch\ min         & \multicolumn{3}{c|}{3 (2 non-electron tracks)}\\\hline
    \wfd\ min  [\gev] & \multicolumn{3}{c|}{2.5}            \\\hline
    \wfd\ max  [\gev] & \multicolumn{2}{c|}{40}   & 60      \\\hline
  \end{tabular}\end{center}
  \caption{The selection cuts applied to each data sample.}
  \label{tab:cuts}
\end{table}
\begin{table}[htb]
  \begin{center}\begin{tabular}{|c|c|c|c|c|}\hline
    \avqsq\ [\gevsq] & sample & luminosity [$\mbox{pb}^{-1}$]  & events & \qsq\ range [\gevsq] \\\hline
    1.9 & LEP1 SW  & 74.6  & 4356 & 1.5--2.5    \\
    3.7 &          &       & 4010 & 2.5--6.0    \\\hline
    8.9 & LEP1 FD  & 97.8  & 1909 & 6.0--12.0   \\
    17.5 &         &       & 1578 & 12.0--30.0  \\\hline
    10.7 & LEP2 SW & 222.9 & 4593 & 7.0--13.0   \\
    17.8 &         &       & 5495 & 13.0--30.0  \\\hline
  \end{tabular}\end{center}
  \caption{The integrated luminosity, number of selected events, 
           and \qsq\ range for each data sample.
           The error on the luminosity is negligible.}
  \label{tab:samples}
\end{table}
%
%
\section{Monte Carlo and background}
\label{sec:MC}
Monte Carlo programs are used to simulate \gsg\ events and to provide 
background estimates. 
All Monte Carlo events are passed through the OPAL detector simulation~\cite{ref:GOPAL} 
and the same reconstruction and analysis chain as real events.

It has been seen in previous studies~\cite{ref:OPAL,ref:ALEPH,ref:L3} 
that Monte Carlo modelling of the hadronic final 
state is a large source of systematic error in the \ft\
measurement. While there is still no 
\gsg\ Monte Carlo generator which describes all the features of
the observed hadronic final state, 
improved Monte Carlo programs have become available 
since the previous OPAL measurements of \ft\
and it is now possible to reject some of the models which do not 
describe the data adequately.

The Monte Carlo generators used to simulate signal events are
HERWIG 5.9~\cite{ref:Herwig},
HERWIG 5.9+\kt~\cite{ref:HKT},
PHOJET 1.05~\cite{ref:Phojet},
and F2GEN~\cite{ref:TWOGEN}.
HERWIG 5.9 is a general purpose Monte Carlo program which includes
deep inelastic electron-photon scattering. 
The HERWIG 5.9+\kt\ version uses a modified
transverse momentum distribution, $k_t$, for the quarks
inside the photon, with the upper limit dynamically (dyn) adjusted according
to the hardest scale in the event, which is of order \qsq.
The distribution was originally tuned for 
photoproduction events at HERA~\cite{ref:HERAKT}.
PHOJET 1.05 simulates hard interactions through perturbative QCD and 
soft interactions through Regge phenomenology and 
thus is expected to provide a more complete picture of \gsg\ 
collisions than HERWIG 5.9. 
It is used here for the first time in an OPAL \ft\ analysis.
In F2GEN, \gsg\ events are 
generated with a pure \qq\ final state.
No QCD effects between the two quarks are 
simulated, and the pointlike mode was used, in which 
the angular distribution of the \qq\ final
state is taken to be the same as for a lepton pair.
The generated luminosities of the Monte Carlo samples were
3--6 times the data luminosity.
 
A sample was generated with each of the Monte Carlo models using
the GRV LO~\cite{ref:GRV} parameterisation of \ft,
taken from the PDFLIB library~\cite{ref:PDFLIB},
as the input structure function. This version
assumes massless charm quarks.
To study the effect of a different input structure 
function on the final state, another  
sample was generated using 
HERWIG 5.9\footnote{The \kt\ version was not used for this test.}
with the SaS1D~\cite{ref:SaS1D} parameterisation of \ft.

All of the Monte Carlo programs except PHOJET 1.05 allow 
generation of events and cross-section calculation according 
to a chosen structure function. PHOJET 1.05
uses an input structure function for the total cross-section 
calculation but always produces the same $x$ and \qsq\
distributions. 
The measurement of \ft\ requires knowledge of the 
structure function used to generate these 
distributions, so the effective internal structure function in PHOJET 1.05 was 
found by comparing the generator level distributions 
to those of HERWIG 5.9+\kt\ with the GRV LO structure function. 
The ratio of the $x$ distributions in the two samples 
for each \qsq\ range was fitted to a polynomial, which, after
multiplying by the GRV LO structure function, 
gave the structure function for the PHOJET 1.05 sample.

The dominant background comes from the reaction
$\epem\rightarrow\epem\gsg\rightarrow\epem\tptm$ 
proceeding via the multiperipheral process shown in Figure~\ref{fig:dis},
with $\epem\rightarrow\epem\gsg\rightarrow\epem\epem$ giving a smaller contribution.
These were simulated using the Vermaseren program~\cite{ref:Vermaseren}.
The process \znhad\ also contributes significantly at LEP1
and was simulated using JETSET~\cite{ref:Jetset}.
 Because the aim is to measure the structure function of the 
 quasi-real photon, the hadronic \gsgs\ events are also treated as background.
 They were generated using PHOJET 1.05 with the virtuality of the 
 target photon restricted to $\psq>$1.0 \gevsq\ (LEP1) 
 or 4.5 \gevsq\ (LEP2).
Other sources of background considered were
\zntau, simulated with KORALZ~\cite{ref:Koralz},
non-multiperipheral four-fermion events with 
\epem\tptm\ and \epem\qq\ final states, 
which were simulated with FERMISV~\cite{ref:FERMI}, 
and \wwhad\ and untagged \grg\ events, which were 
both simulated with PYTHIA~\cite{ref:pythia}. 
The contributions from all these were found to be negligible in all the samples. 
The number of expected signal and
background events for each data sample is shown in Table~\ref{tab:back}.
\begin{table}[htb]
  \begin{center}\begin{tabular}{|c|c|c|c|c|c|}\hline
  \avqsq\ [\gevsq] & signal & \ggtau    & \ggep & \znhads & \gsgshads \\\hline
1.9 & 4237$\pm$66 & 85$\pm$6  & 7$\pm$1 & 10$\pm$2 & 18$\pm$3\\
3.7 & 3834$\pm$64 & 115$\pm$7 & 9$\pm$1 & 18$\pm$3 & 33$\pm$4\\\hline
8.9 & 1749$\pm$45 & 85$\pm$6  & 3$\pm$1 & 42$\pm$6 & 30$\pm$4 \\
17.5 & 1378$\pm$41 & 71$\pm$6  & 4$\pm$1 & 84$\pm$8 & 42$\pm$5 \\\hline
10.7 & 4403$\pm$68 & 135$\pm$6 & 4$\pm$1 & 9$\pm$1  & 25$\pm$2\\
17.8 & 5171$\pm$74 & 220$\pm$7 & 6$\pm$1 & 14$\pm$1 & 45$\pm$2\\\hline
  \end{tabular}\end{center}
  \caption{The measured number of signal events
           (observed events corrected for background) in each data sample,
           and the expected number of background events according to Monte 
           Carlo. Statistical errors only are given.}
  \label{tab:back}
\end{table}

Figures~\ref{91swgen1},~\ref{91fdgen1} and~\ref{189swgen1}
show comparisons between data and  
Monte Carlo distributions for variables relating to the
scattered electrons.
The main differences are in the LEP1 SW data sample,
where the observed cross-section for selected events is
significantly higher than either of the Monte Carlo predictions.
The LEP1 SW data sample is more peaked than the 
HERWIG 5.9+\kt\ prediction at low \ttag,
and there are large discrepancies between the LEP1 SW data sample and
the Monte Carlo samples in the distribution of $\ea/\eb$
(Figure~\ref{91swgen1}d), with the data having 
an excess at the low end of the plot.
However, in this region, the anti-tag distribution is 
more influenced by the hadronic final
state than by the anti-tagged electrons, 
which are almost all above the cut.
 
In the distributions of variables related to the hadronic final 
state (Figures~\ref{91swgen2}--\ref{189sw2dvar}), 
there are large discrepancies.
Figure~\ref{eflows} shows the hadronic energy flow as a function of
pseudorapidity. It can be seen that
HERWIG 5.9 tends to put too little energy in the central 
region of the detector, while F2GEN has a much higher peak
in the central region than the data. 
PHOJET 1.05 gives the best description of the data in these plots.

The large differences between HERWIG 5.9+\kt\ and PHOJET 1.05 in the
\wfd\ distributions
(Figures~\ref{91swgen2}a,~\ref{91fdgen2}a 
and~\ref{189swgen2}a) 
and the \xfd\ distributions 
(Figures~\ref{91swgen2}b,~\ref{91fdgen2}b 
and~\ref{189swgen2}b) 
are mostly due to the fact that PHOJET 1.05 does not use the input
structure function to generate the $x$ distribution, although there are also
differences arising from the final state modelling.
The quantity \xfd\ is calculated
by inserting \wfd\ into equation~\ref{eqn:Xcalc}.
For the number of tracks, \nch, 
in the LEP1 SW sample (Figure~\ref{91swgen2}c),
the data distribution is above the Monte Carlo
distributions at high \nch. 
The Monte Carlo descriptions of \nch\ are closer to the
data in the LEP1 FD and LEP2 SW samples 
(Figures~\ref{91fdgen2}c and ~\ref{189swgen2}c). 
There are significant differences in the distributions of \etout\ 
(Figures~\ref{91swgen2}d,~\ref{91fdgen2}d and~\ref{189swgen2}d on a linear scale and
Figures~\ref{91sw2dvar},~\ref{91fd2dvar} and~\ref{189sw2dvar} on a logarithmic scale and
divided into three bins of \xfd).
\etout\ is the transverse hadronic energy out of the plane 
containing the beam line and the tagged electron:
\begin{equation}
  \etout=\sum_{i} E_{i,\rm t}|\sin\phi_{i}|
  \label{eqn:etout}
\end{equation}
where $i$ runs over all particles in the hadronic final state, 
with transverse energy $E_{i,\rm t}$. The azimuthal angle $\phi_{i}$
is measured from the tagged electron.
HERWIG 5.9 has too few events at large \etout\ and F2GEN has
too many, especially at low values of \xfd.
Also shown in 
Figures~\ref{91sw2dvar},~\ref{91fd2dvar} and~\ref{189sw2dvar}
is \edist, the observed energy in the forward region
divided by the total observed energy. 
Of the four Monte Carlo models, PHOJET 1.05 shows the best  
agreement with the data in the hadronic final state, 
HERWIG 5.9+\kt\ describes the data better than HERWIG 5.9,
and F2GEN gives the worst description of the data. 
For the final results, only PHOJET 1.05 and HERWIG 5.9+\kt\ are used;
however, all four Monte Carlo models are used in studies of the 
unfolding methods.

 None of the Monte Carlo programs discussed above contain
 radiative corrections to the deep inelastic scattering process.
 These are dominated by initial state radiation from the deeply
 inelastically scattered electron.
 The final state radiation is experimentally integrated out due to
 the finite granularity of the calorimeters. The Compton scattering
 process contributes very little, and the radiative corrections
 due to radiation of photons from the electron that produced
 the quasi-real target photon were shown to be very small~\cite{ref:EPACOR}.

 In this analysis the radiative corrections have been evaluated 
 using the RADEG program~\cite{ref:RADEG}, which includes
 initial state radiation and the Compton scattering process.
 The calculations are performed using mixed variables, which means that
 \qsq\ is calculated from electron variables,
 while $W^2$ is calculated from hadronic variables. 
 The value of $x$ is found from $W^2$ and \qsq, 
 exactly as for the experimental analysis.
 The RADEG program allows calculation of differential cross-sections
 in bins of $x$ and \qsq, with or without radiative corrections,
 while applying the experimental restrictions on the electron energy
 and polar angle, the minimum invariant mass 
 and the anti-tag angle.
 The predicted ratio of the differential cross-section for each bin
 of the analysis is used to correct the data, i.e. the measured \ft\
 is multiplied by the ratio of the non-radiative and the radiative
 cross-sections.  
 The radiative corrections reduce the cross-section in the phase space
 of the present analysis. They are largest at low values of $x$. 
 The size of the largest correction to any bin is 14$\%$.
 The radiative corrections are insensitive to the choice
 of \ft\ in the calculation, for example, the difference between the predicted 
 corrections for the GRV and the QPM structure functions is negligible
 in all regions of $x$ and \qsq.

 Several theoretical ansatzes exist for how \ft\ should behave 
 as a function of \psq~\cite{ref:GRVP2,ref:SASP2,ref:DGP2}. 
 They all predict a decrease of \ftxqp\ with \psq\ which is strongest 
 at low values of $x$. 
 This means that applying a correction to obtain an \ft\ which is valid at 
 $\psq=0$ would change the shape of the measured \ft.
 The size of the effect was studied
 based on the GRS~\cite{ref:GRVP2} and SaS1D~\cite{ref:SASP2}  
 parameterisations
 using three values of \psq: $0.01, 0.02$ and $0.03$~\gevsq, and four
 values of \qsq: $1.9, 3.7, 9.5$ and $17.6$~\gevsq. 
 These values reflect the 
 range of \pzm\ values expected in the data, and the 
 \qzm\ values of the data samples.
 The largest suppression observed at $x=0.003$ 
 for $\psq= 0.01/0.02/0.03$~\gevsq\ 
 is around $7/12/17\%$; however, at low $x$ the predictions differ by 
 more than a factor of two.
 Because the distribution of \psq\ in the data is not known and the
 theoretical predictions differ significantly, no correction is applied.
 \par
%
\section{Determination of \boldmath \ft}
\subsection{Unfolding}
Unfolding is a statistical technique which is used in this analysis
to correct the measured distributions for detector effects.
The unfolding problem can be described as follows: A measurement 
$g_j$ is made of a true distribution $f_i$, 
with $j$ and $i$ bins respectively.
The $g_j$ distribution
differs from $f_i$ because of
\begin{itemize}
\item Limited resolution:\\
      Events in a certain bin of the true distribution can be spread
      into several bins in the measured distribution.
      In \gsg\ events
      this can be due to missing energy in the forward region 
      (FD and SW plus the beampipe), 
      as well as the intrinsic
      energy resolution of the subdetectors.
\item Limited acceptance:\\
      Some events fail the cuts and therefore contribute
      only to the true $x$ distribution, and not to the
      measured distribution.
\item Background:\\
      There are also events originating from other reactions which only 
      contribute to the measured distribution.
\end{itemize}
Resolution and acceptance effects can be described in terms
of a response matrix $A_{ij}$. The distributions $g_j$
and $f_i$ are then related by
\begin{equation}
  g_{j}=\sum_{i=1}^{n} A_{ij}f_{i}+b_{j}
  \label{eqn:unfold}
\end{equation}
where $b_{j}$ is the background distribution.
Equation~\ref{eqn:unfold} cannot simply be inverted to find $f_{i}$
because this does not lead to a stable 
solution\footnote{
  See the GURU~\cite{ref:GURU} or RUN~\cite{ref:RUN} documentation for an 
  explanation of this effect.}. 
Instead, after $b_{j}$ is subtracted, $f_{i}$ is estimated 
from a procedure which includes a smoothing (or regularisation) 
of the distribution to reduce statistical fluctuations. 
The response matrix $A_{ij}$ is found from Monte Carlo simulation.
After unfolding, \ft\ is determined 
by reweighting the input structure function of the
Monte Carlo according to the ratio of the unfolded $x$
distribution to the $x$ distribution in the Monte Carlo:
\begin{equation}
  \ftmi = \frac{\langle f(x)\thinspace\ftmc(x) \rangle_i}{N_{{\rm MC},i}}
  \label{eqn:findft}
\end{equation}
where \ftmc\ is the input structure function of the Monte Carlo sample used
for unfolding, $N_{{\rm MC},i}$ is the number of events in the Monte Carlo
sample in the $i$th bin and 
$f(x)$ is the unfolded $x$ distribution, obtained from a quadratic
spline fit to $f_i$, with $f(x)=0$ at the edges
of the acceptance region.
\ftmi\ represents the measured value of the average \ft\ of the events
in bin $i$ of the unfolded distribution, and not the average over
$x$. The distinction is important mainly at low $x$, where $f(x)$
changes most rapidly. 

The main unfolding program used for this analysis was
GURU~\cite{ref:GURU}; the programs RUN~\cite{ref:RUN}
and BAYES~\cite{ref:BAYES} 
were used to check the unfolding.
In previous OPAL \ft\ analyses, only RUN was used.
The programs GURU and RUN are similar in principle,
but have different implementations.
Due to its use of standard matrix techniques,
GURU has a much simpler structure and is easier to modify than RUN,
which uses a maximum likelihood fit with spline functions.
The BAYES program performs unfolding using an iterative method 
based on Bayes' theorem, leaving regularisation to the user
(e.g. a polynomial fit after each iteration except the last). 
The programs GURU, RUN and BAYES were compared by unfolding test
distributions including experimental data. The result of one such 
study, unfolding the LEP1 SW data with HERWIG 5.9+\kt,
using the measured distribution \xfd, 
is shown in Figure~\ref{uf1}a. 
The results from the three programs are generally consistent, though
the BAYES program tends to assign smaller errors than the other two.

The amount of regularisation in the unfolding is set by the number of
degrees of freedom (NDF). It can be seen in 
Figure~\ref{uf1}b that increasing this number 
leads to larger unfolding errors. Reducing the number of
degrees of freedom increases the correlations between the bins.   
The number of degrees of freedom to be used is determined by the GURU program 
from statistical analysis of the data distributions, 
as described in the GURU documentation~\cite{ref:GURU}.

The unfolding of the variable $x$ is done on a logarithmic scale in 
order to study the low-$x$ region in more detail.
Several improvements to the unfolding procedure used in previous 
OPAL analyses have been implemented. These are described in the 
following sections.
\subsection{Reconstruction of $\boldmath W$}
The visible hadronic mass squared is defined by
\begin{equation}
  \wfdsq = (\sum_{i} E_i)^2-(\sum_{i} p_{i,x})^2-(\sum_{i} p_{i,y})^2-(\sum_{i} p_{i,z})^2
  \label{eqn:wdef}
\end{equation}
where $i$ runs over all tracks and clusters in the hadronic final state, 
with energy $E_i$ and momentum vector $(p_{i,x},p_{i,y},p_{i,z})$.
Because hadronic showers are not well contained in the 
electromagnetic calorimeters, and some energy is lost in the beampipe,
the measured energy in the forward region is less than the true energy. 
It is possible to improve the reconstruction of \wfd\ 
by including kinematic information from the tagged electron~\cite{ref:JaquetBlondel}.
\wfdsq\ can also be written as
\begin{equation}
  \wfdsq = (\sum_{i} p_{i+})(\sum_{i} p_{i-}) - (\sum_{i} p_{i,\rm t})^2
  \label{eqn:jb1}
\end{equation}
where $p_{i\pm}=E_i \pm p_{i,z}$ and $p_{i,\rm t}=\sqrt{{p_{i,x}}^2+{p_{i,y}}^2}$.
Using conservation of energy and momentum, 
and assuming that the untagged electron travels along the 
beam direction, the sums over $p_{i+}$ and $p_{i,\rm t}$ can be 
replaced by electron variables,
leading to
\begin{equation}
  \wrecsq = (p_{\rm beam+}-p_{\rm tag+})(\sum_{i} p_{i-}) - (p_{\rm tag,t})^2
  \label{eqn:jb2}
\end{equation}
where $p_{\rm beam+}$ and $p_{\rm tag+}$ are calculated for the
tagged electron before and after scattering, respectively, and 
$p_{\rm tag,t}$ is calculated according to the definition of
$p_{i,\rm t}$, above, for the tagged electron. 
When using Equation~\ref{eqn:jb2} instead of Equation~\ref{eqn:jb1} 
to evaluate the measured hadronic mass of each event, the hadronic energy
enters only through the $p_{i-}$ term.  This is an advantage because
the leptonic energy resolution is usually better than the hadronic energy
resolution. The (reconstructed) variable formed in
this way is called $W_{\rm rec}$. The corresponding measurement
of $x$ is called \xrec.
The improvement in the resolution when using \wrec\ instead of \wfd\
depends on how well the hadronic system is measured, and is therefore 
both detector dependent and model dependent.

Even after the above technique has been applied, the value of $W_{\rm rec}$ is
still generally smaller than the true value. This is mainly due to
energy losses in the forward calorimeters, in which
only about 40\% of the hadronic energy is observed. 
This makes the measurement
very dependent on the angular distribution of
the hadrons in the final state.
In an attempt to make the energy response of the detector more
uniform and to reduce the systematic error due to the uncertainty in
the Monte Carlo modelling, a new (corrected) variable is
formed: $W_{\rm cor}$, along with the corresponding $x$ measurement
\xcor, in which the contribution from the forward 
region has been scaled by a factor of 2.5. This factor was obtained by
comparing the generated and measured energy in the forward region in
Monte Carlo events. The uncertainty of this
factor has been taken into account
in the evaluation of the systematic errors.
\begin{eqnarray}
  \wcorsq &=& (p_{\rm beam+}-p_{\rm tag+})(\sum_{i} p'_{i-}) - 
  (p_{\rm tag,t})^2
  \label{eqn:wcor} \\
  p'_{i-} &=& \left\{ \begin{array}{ll}
               2.5p_{i-}  & \mbox{ for clusters observed in SW or FD} \\
               p_{i-}   & \mbox{ otherwise} \\
              \end{array} \right. \nonumber
\end{eqnarray}
Figure~\ref{wcor}a shows 
the correlation between the generated $W$ 
and the three measured quantities $W_{\rm vis}$,
$W_{\rm rec}$ and $W_{\rm cor}$ for the 
LEP2 SW sample generated with HERWIG 5.9+\kt. 
The spread of $W_{\rm cor}$ in $W$ 
is larger than that of the other two variables, because in 
$W_{\rm cor}$ more weight is given to the
forward region, where the energy resolution is worse than in the 
central region.
In general, \wcor\ is still lower than $W$, mainly because of 
energy lost in the beampipe.
Figure~\ref{wcor}b shows the correlation between the generated
energy in the forward region and the scaled observed
energy in that region, $E_{\rm cor}$.

The three measured variables \xfd, \xrec\ and \xcor\ can be used to
unfold the true variable $x$. The difference between the results is 
model dependent and generally small when using Monte Carlo models 
that already give a good description of the hadronic final state
(Figure~\ref{uf1}c).
\subsection{Two dimensional unfolding}
The GURU program can also be used to perform unfolding 
in two dimensions. 
As with the attempts to improve the $W$ reconstruction described above, 
the motivation is to reduce the dependence of the unfolding 
on a particular Monte Carlo model. There is information in
every event about the angular distribution of energy in the detector,
but it is not fully exploited if only $x$ is used in the unfolding procedure. 
Including another variable in a second unfolding dimension makes
more direct use of this information.

Two variables were considered as possible second unfolding variables.
They were chosen because they are very sensitive to the angular
distribution of the hadrons in the final state:
\begin{itemize}
  \item \etdist, the transverse hadronic energy out of the plane 
        containing the beam line and the tagged electron 
        (see Equation~\ref{eqn:etout}),
        divided by the total observed energy. 
  \item \edist, the observed energy in the forward calorimeters
        divided by the total observed energy. 
\end{itemize}
These variables are shown in Figures~\ref{91sw2dvar}--\ref{189sw2dvar}. 
To test the two dimensional unfolding procedure, a random number 
was used as the second variable. The results are shown in Figure~\ref{uf1}d.
They are consistent with
one dimensional unfolding, which is as expected since
there is no extra information in a random number. 
\subsection{Testing the unfolding methods}
\label{sec:test}
To find which of the unfolding methods was the most reliable, 
OPAL data was unfolded with four Monte Carlo programs 
HERWIG 5.9, HERWIG 5.9+\kt\, PHOJET 1.05 and F2GEN.
Despite not giving good descriptions of the data, HERWIG 5.9 and F2GEN
were included in this study to investigate the effectiveness of the
new techniques using extreme models.
The best methods are considered to be those which give the 
smallest difference between the unfolded results with the 
four Monte Carlo models. The quantity compared is \chmodels, defined by
\begin{equation}
\chmodels=\frac{1}{4}\sum_{\rm models}
       \sum_{i}\left(\frac{\ft_i-\langle\ft_i\rangle}{\sigma_i}\right)^2
\label{eqn:chi2}
\end{equation}
where $\ft_i$ is the value of the unfolded result in the $i$th bin, 
$\langle\ft_i\rangle$ is the average of the results from all four
Monte Carlo models and
$\sigma_i$ is the statistical error for $\ft_i$.
The values of \chmodels\ are shown in Table~\ref{tab:ufchi} and
the results of the unfolding are shown in
Figures~\ref{91swuf2}--\ref{189swuf2}.
The \chmodels\ values show how large the differences between the models
are, compared to the statistical error.
In general, the lowest values of \chmodels\ were obtained using two-dimensional
unfolding with \xcor\ as the first variable and \etdist\ as the second
variable, which consequently is used as the standard unfolding method
for the results.
The \chmodels\ values are usually smaller for the LEP1 FD sample 
than the other two. This is partly due to the smaller number of bins.
Additionally, the lower statistics in the LEP1 FD sample mean that any
difference between the Monte Carlo programs would be less significant.

With both two dimensional unfolding
and with one dimensional unfolding using \xcor\ as the unfolding variable, 
the spread between the results with the different models is reduced compared to
one dimensional unfolding results using \xfd, and the
agreement between F2GEN and the other models is especially improved.
Using different unfolding methods with the 
HERWIG 5.9+\kt\ and PHOJET 1.05 samples
makes less difference than with the other two Monte Carlo samples,
which is as expected for models which give a
better description of the data.

As a final test of the unfolding method, several samples of Monte Carlo
events generated using HERWIG 5.9 with the SaS1D structure function
were unfolded using HERWIG 5.9 with the GRV LO structure
function. The results are shown in Figure~\ref{sastest}.
 The measured structure function agrees with the input structure function
 within the statistical precision of the Monte Carlo events generated using
 the SaS1D structure function.

 The effect of taking into account the shape of $f(x)$ within each bin of $x$,
 as in Equation~\ref{eqn:findft}, compared to simply extracting 
 \ft\ as the average input structure function of the Monte Carlo sample
 times the ratio of the number of unfolded events and the number of
 Monte Carlo events in each bin, was also investigated with these samples.
 The effect is largest in the lowest $x$ bins for LEP2 SW, where the
 decrease in the value of \ft\ when taking into account the shape within
 the bin is a little less than the statistical errors shown in 
 Figure~\ref{sastest}.

\begin{table}[htb]
  \begin{center}\begin{tabular}{|c|c|c|c|c|c|c|c|}\cline{3-8}
   \multicolumn{2}{c|}{} & \multicolumn{6}{c|}{\chmodels}\\\cline{3-8}
    \multicolumn{2}{c|}{}& \multicolumn{2}{c|}{1D} 
   & \multicolumn{2}{c|}{2D, \etdist}
   & \multicolumn{2}{c|}{2D, \edist} \\\hline
  \avqsq\ [\gevsq]& sample & \xfd  & \xcor & \xrec & \xcor & \xrec & \xcor\\\hline
 1.9 & LEP1 SW & 66.8 & 29.1 & 22.1 & 9.6 & 24.7 & 26.9 \\
 3.7 & & 53.1 & 26.1 & 13.5 & 8.1 & 19.5 & 16.8 \\ \hline
 8.9 & LEP1 FD & 15.4 & 6.8 & 7.1 & 3.4 & 10.7 & 8.7 \\
 17.5 & & 8.5 & 4.5 & 8.8 & 12.4 & 4 & 4.1 \\ \hline
 10.7 & LEP2 SW & 62.5 & 22.4 & 18.7 & 6.8 & 20.6 & 16.8 \\
 17.8 & & 57.2 & 18.3 & 13 & 8.4 & 57.7 & 15.4 \\ \hline
  \end{tabular}\end{center}
  \caption{\chmodels, as defined in Equation~\ref{eqn:chi2}, 
           for different unfolding methods. 
           The number of bins in $x$ was 4 for the LEP1 SW and LEP2 SW
           samples, and 3 for the LEP1 FD samples.  
          }
  \label{tab:ufchi}
\end{table}
%
%
\section{Results}
The photon structure function \ft\ is measured by 
unfolding each data sample in bins of $\mbox{log}(x)$.
Each OPAL data sample is divided into two ranges of \qsq\ containing
approximately equal numbers of events.
The ranges correspond to
\qzm\ values of 1.9 and 3.7~\gevsq\ for the LEP1 SW sample, 
and 8.9 (10.7) and 17.5 (17.8)~\gevsq\ for the LEP1 FD (LEP2 SW)
sample, where the average \qsq\ values have been obtained from the
data.
\ft\ is unfolded to a given \qzm\ value and does not
correspond to an average in each bin of \qsq.

The results are listed in Table~\ref{tab:resxq}. 
The quoted values were measured as the average \ftn\
in each bin of $x$ weighted by the unfolded $x$ distribution, 
according to Equation~\ref{eqn:findft}, then corrected to the log centre 
of each bin, except for the highest $x$ bins where the log centre of that
portion of the bin below the charm threshold for $m_c=1.5$~\gev\ was used.
The bin-centre corrections are the average of the GRV LO and
SaS1D predictions for the correction from the average \ft\
over the bin to the value of \ft\ at the nominal $x$ position.

The results are also corrected for radiative effects. The radiative
corrections were calculated using the RADEG~\cite{ref:RADEG} program
and are listed in Table~\ref{tab:radcor}, along with the bin-centre 
corrections.
The statistical correlations between bins are
shown in Table~\ref{tab:corr}.

The central value  of \ft\ in each $x$ bin is the average of the 
data unfolded with HERWIG 5.9+\kt\ and PHOJET, using two dimensional 
unfolding with \xcor\ as the first variable and 
\etdist\ as the second variable.
Standard HERWIG 5.9 and F2GEN were not used as 
they are not in acceptable agreement with the data.
The systematic errors are evaluated by repeating the unfolding
with one parameter varied at a time and finding the shift in the result.
The systematic errors are combined by adding all of the individual
contributions in quadrature. The summing is done separately for
positive and negative errors.
The systematic effects considered are listed below.
\begin{itemize}
\item Monte Carlo modelling:\\    
      The quoted result is the average of the results obtained using
      HERWIG 5.9+\kt\ and PHOJET 1.05. The errors are symmetrical
      and equal to half the difference between the results 
      with the two Monte Carlo programs. 
\item Unfolding method:\\
      The unfolding was repeated with \edist\ as the 
      second variable, instead of \etdist.
\item Unfolding parameters:\\
      The number of bins used for the measured variable
      can be different from the number of bins used for the true variable
      (though it should be at least as large).
      The standard result has 6 bins in the measured variable. 
      This was increased to 8 to estimate the
      systematic effects of unfolding.
\item $W$ reconstruction:\\
      The weighting applied to the energy in the forward calorimeters 
      was varied between 2.0 and 3.0.
      This check allows for uncertainty in the treatment of forward
      energy.
\item Variations of cuts:\\
      The composition of the selected events was varied by
      changing the cuts one at a time. The size of the variations
      reflect the resolution of the measured variables and
      are sufficiently small not to change the average \qsq\
      of the sample significantly, or remove too many events from
      any single $x$ bin.
      The variations of the cuts are listed in Table~\ref{tab:cutvar}.
\item Off-momentum electrons:\\
      There is a possible contamination
      of off-momentum background in one region of azimuthal angle $\ph$ 
      in the 183~\gev\
      data, so as a precaution this $\ph$ region was removed in the
      measurement of \ft, and the difference from the result with the 
      $\ph$ region left in was included as a systematic error.
\item Calibration of the tagging detectors (FD and SW):\\
      The energy of the tagged electron in the Monte Carlo samples was 
      scaled by $\pm 1\%$, to allow for
      uncertainty in the simulation of the detectors. 
      The size of the variation was motivated 
      by a comparison of the $\etag/\eb$ distributions in data and Monte Carlo.
\item Measurement of the hadronic energy:\\
      The main uncertainty is in the calibration of the 
      electromagnetic calorimeters (excluding the forward region which 
      is dealt with separately). This was varied by $\pm$3\% 
      in the Monte Carlo samples.
      The track quality criteria were also varied, 
      but the effects were negligible.
\item Simulation of background:\\
      The hadronic background events at LEP1 stem from the production 
      of photons or light mesons with a large fraction of the beam
      energy. The mesons fake an electron tag due to their decays into
      photons.
      The cross-section for these events was measured by OPAL with an
      accuracy of about 50\%, and found to be consistent with the 
      JETSET prediction~\cite{ref:mesons}.
      The normalisation of the simulated background was 
      therefore varied by $\pm$50\%.
\item Bin-centre correction:\\
      The corrections depends on the shape of \ft\ in each bin. The
      average of the corrections based on GRV LO and SaS 1D was used, 
      as these parameterisations are the closest to the data.
      The error is half the difference between the GRV LO and SaS 1D
      corrections, and is symmetric.  
\end{itemize}

Because the estimation of each source of systematic error has some 
statistical fluctuation due to the changing distributions 
(e.g. when a cut is varied), the quadratic sum of all individual sources is an 
overestimate of the total systematic error.
The expected statistical component of each source of systematic error
that changes the data sample
is determined by unfolding 8 Monte Carlo samples, each about 
the same size as the data sample. 

The corrected systematic error on the $i$th bin from source $k$ is then 
\begin{eqnarray}
   (\Delta' f_{i,k})^2&=&(\Delta f_{i,k})^2-\sigma_{\Delta f_{i,k}}^2
    \mbox{ for }\Delta f_{i,k}^2>\sigma_{\Delta f_{i,k}}^2 \nonumber \\
   \Delta' f_{i,k}&=&0 \mbox{ otherwise}
\end{eqnarray}
where $\Delta f_{i,k}$ is the shift on the $i$th bin in the data 
when changing a parameter $k$, and $\sigma_{\Delta f_{i,k}}$ is the 
expected statistical component of the shift, which
is approximated by the statistical spread in the systematic 
error estimates for source $k$ in the 8 Monte Carlo samples. 
In a few cases this was larger than the 
statistical error in the data. In these cases $\sigma_{\Delta f_{i,k}}$
was set to the statistical error in the data, in order not to hide possibly
significant systematic errors.
This procedure means that for the individual systematic errors either the 
expected statistical component is subtracted, or the error is 
set to zero if the observed shift in the data is consistent with a statistical 
fluctuation as predicted by the 8 Monte Carlo samples.

The total systematic error is the quadratic sum of all 
the individual contributions $\Delta' f_{i,k}$ from the above sources.
The sum is made separately for deviations above and 
below the standard result.
The systematic errors from each source as a function of $x$ and \qsq\ 
are listed in Table~9.

The results are shown along with previous OPAL
measurements of \ft~\cite{ref:OPAL} in 
Figures~\ref{result1opal} and~\ref{result2opal}.
The overlapping results from the LEP1 FD and LEP2 SW samples are in
good agreement.
The measurements of \ft\ using LEP1 data for
$\qzm=1.9$~\gevsq\ and $\qzm=3.8$~\gevsq\ 
are lower than the previous OPAL LEP1 results, which were
unfolded using HERWIG 5.8d. Repeating the unfolding with 
HERWIG 5.8d gives results which are consistent with the old
analysis, but with better precision. 
The HERWIG 5.8d Monte Carlo model has now been replaced by 
HERWIG 5.9+\kt, which gives a better description of the data.
The results in the four higher \qsq\ bins are consistent with 
previous OPAL measurements, and have smaller errors.
The previous LEP1 results with electrons tagged in SW or FD 
are superseded by the present analysis.

In Figures~\ref{result1others} and~\ref{result2others} the results are
compared to measurements of \ft\ from other experiments:
TPC/2$\gamma$~\cite{ref:TPC},
PLUTO~\cite{ref:PLUTO}, 
TOPAZ~\cite{ref:TOPAZ},
ALEPH~\cite{ref:ALEPH}, 
DELPHI~\cite{ref:DELPHI} and
L3~\cite{ref:L3}.
 Also shown in Figures~\ref{result1others} and~\ref{result2others}
 are the GRV LO, SaS1D and 
 WHIT1~\cite{ref:WHIT}\footnote{The starting scale of the evolution for the
 WHIT parameterisation is 4~\gevsq; consequently it can only be compared
 to the measurements above $\qzm=4$~\gevsq.} 
 parameterisations of \ft, 
 and the naive quark-parton model (QPM).
 The QPM prediction, which only models the
 point-like component of \ft, is calculated for four active flavours with
 masses of 0.2~\gev\ for light quarks and 1.5~\gev\ for charm quarks.\par
 The previous results are found to be generally consistent with the 
 new OPAL measurements. 
 The largest differences are from the older measurement
 by TPC/2$\gamma$ at $\qzm=2.8$~\gevsq, which suggests
 a different shape of \ft\ to all the other measurements.
 The L3 results obtained at $\qzm = 5.0$~\gevsq\ 
 are consistently higher than the OPAL measurements.
 However, because the L3 points are highly correlated,
 due to the finer binning in $x$, 
 the discrepancy looks stronger than it actually is.
 \par
 Due to large spread in the theoretical predictions for the \psq\ suppression
 of \ft\ (as discussed in Section~\ref{sec:MC}) the OPAL measurements are
 not corrected for this effect. 
 Since the data contain a \psq\ suppression, which is not included
 in the Monte Carlo simulation, applying the correction would
 lead to an $x$ dependent increase of the measured \ft.
 This should be taken into account when comparing to the parameterisations
 of \ft, which are all shown for $\psq=0$.
 In general the shape of the GRV LO parameterisation is consistent 
 with the OPAL data in all the accessible $x$ and \qsq\ regions.
 The normalisation is also consistent with the data, except
 at the lowest scale, $\qzm=1.9$~\gevsq, where GRV is too low.
 The SaS1D LO prediction shows a slower evolution with \qsq\ than the GRV
 prediction. At low \qsq\ it is slightly above GRV LO, whereas at the largest
 \qsq\ values shown, it falls below GRV LO. Within the precision of the 
 OPAL measurement, the description of the data by SaS1D LO is of similar
 quality to GRV LO.
 In the region of applicability, the WHIT1 prediction is higher than
 the OPAL data and flatter than the other predictions, 
 though the shape is still consistent with the data.
 \par
 The hadron-like component is predicted to dominate at low values of $x$.
 For $x<0.1$ the naive quark-parton model is not able to describe the data, 
 indicating that the photon must contain a significant hadron-like
 component at low $x$.

\begin{table}[tb]
  \begin{center}\begin{tabular}{|c|c|c|c|}\hline
    cut               & LEP1 SW   & LEP1 FD         & LEP2 SW \\\hline
    \etag/\eb\ min    & \multicolumn{2}{c|}{$\pm$0.025 (0.75)} & $\pm$0.025 (0.775) \\\hline
    \ttag\ min [mrad] & +2 (27)   & +2 (60)         & +2 (33.25) \\\hline
    \ttag\ max [mrad] & $-$2 (55) & $-$2 (120)      & $-$2 (55) \\\hline
    \ea/\eb\ max      & \multicolumn{3}{c|}{$\pm$0.05 (0.25)} \\\hline
    \wfd\ min  [\gev] & \multicolumn{3}{c|}{+1.0 (2.5)}       \\\hline
    \wfd\ max  [\gev] & \multicolumn{2}{c|}{$\pm$5 (40)} & $\pm$5 (60) \\\hline
  \end{tabular}\end{center}
  \caption{Systematic variations in the cuts. The standard cuts are
           given in brackets after the variations.}
  \label{tab:cutvar}
\end{table}
\renewcommand{\arraystretch}{1.30}
\begin{table}[htb]
\begin{center}
\begin{tabular}{|c|c|c|c|c|c|}\hline
 \qzm\ [\gevsq] & sample & bin & $x$ range & $x$ & \ftn                 \\\hline

 1.9 & LEP1 SW
 & I & $ 0.0006 < x < 0.0028 $ & 0.0012 & \Z{0.269}{0.027}{0.018}{0.034} \\
 &
 & II & $ 0.0028 < x < 0.0143 $ & 0.0063 & \Z{0.177}{0.009}{0.017}{0.014} \\
 &
 & III & $ 0.0143 < x < 0.0724 $ & 0.0322 & \Z{0.179}{0.007}{0.007}{0.006} \\
 &
 & IV & $ 0.0724 < x < 0.3679 $ & 0.1124 & \Z{0.227}{0.010}{0.012}{0.012} \\
\hline 
 3.7 & LEP1 SW
 & I & $ 0.0015 < x < 0.0067 $ & 0.0032 & \Z{0.269}{0.033}{0.047}{0.033} \\
 &
 & II & $ 0.0067 < x < 0.0302 $ & 0.0143 & \Z{0.232}{0.013}{0.023}{0.021} \\
 &
 & III & $ 0.0302 < x < 0.1353 $ & 0.0639 & \Z{0.259}{0.010}{0.006}{0.013} \\
 &
 & IV & $ 0.1353 < x < 0.6065 $ & 0.1986 & \Z{0.296}{0.014}{0.029}{0.022} \\
\hline 
 8.9 & LEP1 FD
 & I & $ 0.0111 < x < 0.0498 $ & 0.0235 & \Z{0.221}{0.017}{0.030}{0.026} \\
 &
 & II & $ 0.0498 < x < 0.2231 $ & 0.1054 & \Z{0.308}{0.014}{0.011}{0.012} \\
 &
 & III & $ 0.2231 < x < 0.8187 $ & 0.3331 & \Z{0.379}{0.022}{0.017}{0.015} \\
\hline 
 10.7 & LEP2 SW
 & I & $ 0.0009 < x < 0.0050 $ & 0.0021 & \Z{0.362}{0.045}{0.058}{0.039} \\
 &
 & II & $ 0.0050 < x < 0.0273 $ & 0.0117 & \Z{0.263}{0.015}{0.032}{0.030} \\
 &
 & III & $ 0.0273 < x < 0.1496 $ & 0.0639 & \Z{0.275}{0.011}{0.029}{0.030} \\
 &
 & IV & $ 0.1496 < x < 0.8187 $ & 0.3143 & \Z{0.351}{0.012}{0.025}{0.016} \\
\hline 
 17.5 & LEP1 FD
 & I & $ 0.0235 < x < 0.0821 $ & 0.0439 & \Z{0.273}{0.028}{0.032}{0.039} \\
 &
 & II & $ 0.0821 < x < 0.2865 $ & 0.1534 & \Z{0.375}{0.023}{0.020}{0.013} \\
 &
 & III & $ 0.2865 < x < 0.9048 $ & 0.3945 & \Z{0.501}{0.027}{0.027}{0.019} \\
\hline
 17.8 & LEP2 SW
 & I & $ 0.0015 < x < 0.0074 $ & 0.0033 & \Z{0.428}{0.061}{0.055}{0.071} \\
 &
 & II & $ 0.0074 < x < 0.0369 $ & 0.0166 & \Z{0.295}{0.019}{0.033}{0.020} \\
 &
 & III & $ 0.0369 < x < 0.1827 $ & 0.0821 & \Z{0.336}{0.013}{0.041}{0.042} \\
 &
 & IV & $ 0.1827 < x < 0.9048 $ & 0.3483 & \Z{0.430}{0.013}{0.032}{0.025} \\
\hline

\end{tabular}
\caption{Results for \ftn\ as a function of $x$ for four active flavours 
         in bins of \qsq. The first errors are statistical
         and the second systematic.
         The structure function was unfolded in bins defined by the
         $x$ ranges and corrected to the $x$ values given.}
\label{tab:resxq}
\end{center}\end{table}
%
\section{Conclusions}
The photon structure function \ft\ has been measured using
deep inelastic electron-photon scattering events
recorded by the OPAL detector during the years 1993--1995,
1997 and 1998, at \epem\ centre-of-mass energies of 91~\gev\ (LEP1), 
and 183--189~\gev\ (LEP2).
\ft\ has been measured as a function of $x$ 
to the lowest attainable $x$ values,
in six ranges of \qsq\ (including two overlapping pairs) 
corresponding to average \qsq\ values of 1.9, 3.7 \gevsq\ for LEP1 SW, and
8.9 (10.7), 17.5 (17.8) \gevsq\ for LEP1 FD (LEP2 SW).
In previous OPAL studies of the photon structure function, 
it became clear that a large source of uncertainty in the
measurement came from the Monte Carlo modelling 
of the hadronic final state of deep inelastic 
electron-photon scattering events.
Since then,
improved Monte Carlo models have become available.
In a comparison of the energy flows and \etout\ distributions,
which are very sensitive to the modelling of the hadronic
final state, these new models, HERWIG 5.9+\kt\ and PHOJET 1.05,
give better descriptions of OPAL data 
than the HERWIG 5.9 and F2GEN programs.
Consequently the latter two programs have not been used for the 
\ft\ measurement.
Previous OPAL measurements of \ft\ using LEP1 data 
with electrons tagged in SW or FD 
are superseded by this analysis.

To further reduce the Monte Carlo modelling error,
two dimensional unfolding has been introduced, using 
\etdist\ as a second unfolding variable. 
Also, the reconstruction of the invariant mass
of the hadronic final state has been improved by including
information from the deeply inelastically scattered electron, and by
scaling the energy observed in the forward calorimeters to partially
compensate for energy losses.
Monte Carlo modelling of the final state is still a significant source
of systematic error, but it no longer dominates all other
sources. 
The total systematic errors are of comparable size to the 
statistical errors.

 Although the precision of the measurement at low $x$ has been
 considerably improved it is still insufficient to determine
 whether or not there is a rise in \ft\ in that region.
 \par
 The GRV LO and SaS1D parameterisations are generally consistent with the
 OPAL data in all the accessible $x$ and \qsq\ regions, with the exception of
 the measurement at the lowest scale, $\qzm=1.9$~\gevsq, where GRV is too low.
 In contrast, the naive quark-parton model is not able to describe the
 data for $x<0.1$. These results show that the photon must contain a 
 significant hadron-like component at low $x$. 
%
\appendix
\par
\section*{Acknowledgements}
 We particularly wish to thank the SL Division for the efficient operation
 of the LEP accelerator at all energies
 and for their continuing close cooperation with
 our experimental group.  We thank our colleagues from CEA, DAPNIA/SPP,
 CE-Saclay for their efforts over the years on the time-of-flight and trigger
 systems which we continue to use.  In addition to the support staff at our own
 institutions we are pleased to acknowledge the  \\
 Department of Energy, USA, \\
 National Science Foundation, USA, \\
 Particle Physics and Astronomy Research Council, UK, \\
 Natural Sciences and Engineering Research Council, Canada, \\
 Israel Science Foundation, administered by the Israel
 Academy of Science and Humanities, \\
 Minerva Gesellschaft, \\
 Benoziyo Center for High Energy Physics,\\
 Japanese Ministry of Education, Science and Culture (the
 Monbusho) and a grant under the Monbusho International
 Science Research Program,\\
 Japanese Society for the Promotion of Science (JSPS),\\
 German Israeli Bi-national Science Foundation (GIF), \\
 Bundesministerium f\"ur Bildung und Forschung, Germany, \\
 National Research Council of Canada, \\
 Research Corporation, USA,\\
 Hungarian Foundation for Scientific Research, OTKA T-029328, 
 T023793 and OTKA F-023259.\\
%
%



\renewcommand{\arraystretch}{1.30}
\begin{table}[htb]
\begin{center}
\begin{tabular}{|c|c|c|c|c|c|c|}\hline
 \qzm\ [\gevsq] & sample & bin & $x$ range & $x$ & radiative  & bin-centre \\
                &        &     &           &     & correction & correction \\\hline

 1.9 & LEP1 SW
 & I & $ 0.0006 < x < 0.0028 $ & 0.0012 & -12.7 & -4.2 \\
 &
 & II & $ 0.0028 < x < 0.0143 $ & 0.0063 & -9.0 & 0.4 \\
 &
 & III & $ 0.0143 < x < 0.0724 $ & 0.0321 & -7.1 & 1.8 \\
 &
 & IV & $ 0.0724 < x < 0.3679 $ & 0.1124 & -6.0 & 4.7 \\
\hline 
 3.7 & LEP1 SW
 & I & $ 0.0015 < x < 0.0067 $ & 0.0032 & -11.8 & -5.0 \\
 &
 & II & $ 0.0067 < x < 0.0302 $ & 0.0143 & -8.9 & 0.6 \\
 &
 & III & $ 0.0302 < x < 0.1353 $ & 0.0639 & -7.3 & 1.9 \\
 &
 & IV & $ 0.1353 < x < 0.6065 $ & 0.1986 & -6.5 & 1.6 \\
\hline 
 8.9 & LEP1 FD
 & I & $ 0.0111 < x < 0.0498 $ & 0.0235 & -7.7 & 0.9 \\
 &
 & II & $ 0.0498 < x < 0.2231 $ & 0.1054 & -6.3 & 2.4 \\
 &
 & III & $ 0.2231 < x < 0.8187 $ & 0.3331 & -4.1 & -0.9 \\
\hline 
 10.7 & LEP2 SW
 & I & $ 0.0009 < x < 0.0050 $ & 0.0021 & -12.5 & -8.7 \\
 &
 & II & $ 0.0050 < x < 0.0273 $ & 0.0117 & -7.3 & -0.5 \\
 &
 & III & $ 0.0273 < x < 0.1496 $ & 0.0639 & -4.4 & 3.2 \\
 &
 & IV & $ 0.1496 < x < 0.8187 $ & 0.3143 & -2.2 & -1.0 \\
\hline 
 17.5 & LEP1 FD
 & I & $ 0.0235 < x < 0.0821 $ & 0.0439 & -9.4 & 2.1 \\
 &
 & II & $ 0.0821 < x < 0.2865 $ & 0.1534 & -7.9 & 2.5 \\
 &
 & III & $ 0.2865 < x < 0.9048 $ & 0.3945 & -6.5 & 0.0 \\
\hline 
 17.8 & LEP2 SW
 & I & $ 0.0015 < x < 0.0074 $ & 0.0033 & -13.6 & -8.2 \\
 &
 & II & $ 0.0074 < x < 0.0369 $ & 0.0166 & -9.9 & -0.5 \\
 &
 & III & $ 0.0369 < x < 0.1827 $ & 0.0821 & -8.4 & 3.7 \\
 &
 & IV & $ 0.1827 < x < 0.9048 $ & 0.3483 & -7.3 & -0.4 \\
\hline

\end{tabular}
\caption{Corrections to the result
         as a function of $x$ in bins of \qsq, 
         as a percentage of the non-corrected \ft.
         The radiative corrections were predicted by RADEG~\cite{ref:RADEG}. 
         The bin-centre corrections are the average of the GRV LO and
         SaS1D predictions for the correction from the average \ft\
         over the bin to the value of \ft\ at the nominal $x$ position.
         The $x$ positions are at the log centre
         of the bins, except for the highest $x$ bins, where they are
         at the the log centre of that portion of the bin below the 
         charm threshold for $m_c=1.5$~\gev.}
\label{tab:radcor}
\end{center}\end{table}


\begin{table}[p]
  \begin{center}\begin{tabbing}
  
  \hspace{2cm}\= \avqsq=1.9 \gevsq\ (LEP1 SW)
  \hspace{2cm}\= \avqsq=3.7 \gevsq\ (LEP1 SW)\\
  \>
  \begin{tabular}{|c|c|c|c|c|}\hline
        & I     & II    & III   & IV   \\\hline
    I   &  1.00 &       &       &      \\\hline
    II  & -0.28 &  1.00 &       &      \\\hline
    III &  0.03 & -0.35 &  1.00 &      \\\hline
    IV  &  0.01 &  0.10 & -0.48 & 1.00 \\\hline
  \end{tabular}
  \>
  \begin{tabular}{|c|c|c|c|c|}\hline
        & I     & II    & III   & IV   \\\hline
    I   &  1.00 &       &       &      \\\hline
    II  & -0.27 &  1.00 &       &      \\\hline
    III &  0.02 & -0.34 &  1.00 &      \\\hline
    IV  &  0.01 &  0.10 & -0.52 & 1.00 \\\hline
  \end{tabular}\\

  \vspace{1cm}\\

  \>\avqsq=8.9 \gevsq\ (LEP1 FD) 
  \>\avqsq=10.7 \gevsq\ (LEP2 SW) \\
  \>
  \begin{tabular}{|c|c|c|c|}\hline
        & I     & II    & III  \\\hline
    I   &  1.00 &       &      \\\hline
    II  & -0.02 &  1.00 &      \\\hline
    III & -0.12 & -0.31 & 1.00 \\\hline
  \end{tabular}
  \>
  \begin{tabular}{|c|c|c|c|c|}\hline
        & I     & II    & III   & IV \\\hline
    I   &  1.00 &       &       &   \\\hline
    II  & -0.32 &  1.00 &       &   \\\hline
    III &  0.04 & -0.29 &  1.00 &   \\\hline
    IV  &  0.00 &  0.06 & -0.36 &  1.00  \\\hline
  \end{tabular}\\

  \vspace{1cm}\\

  \>\avqsq=17.5 \gevsq\ (LEP1 FD)
  \>\avqsq=17.8 \gevsq\ (LEP2 SW) \\
  \>
  \begin{tabular}{|c|c|c|c|}\hline
        & I     & II    & III  \\\hline
    I   &  1.00 &       &      \\\hline
    II  & -0.09 &  1.00 &      \\\hline
    III & -0.09 & -0.35 & 1.00 \\\hline
  \end{tabular}
  \>
  \begin{tabular}{|c|c|c|c|c|}\hline
        & I     & II    & III   & IV   \\\hline
    I   &  1.00 &  &  &   \\\hline
    II  & -0.35 &  1.00 &  &   \\\hline
    III &  0.05 & -0.33 &  1.00 &   \\\hline
    IV  & -0.01 &  0.09 & -0.41 &  1.00  \\\hline
  \end{tabular}\\

  \end{tabbing}\end{center}
  \caption{Statistical correlations between bins for each sample.
           The numerals refer to the bins listed in Table~\ref{tab:resxq}.
          }
  \label{tab:corr}
\end{table}

\clearpage


\tabcolsep 3pt
\begin{table}[p]
\vspace{-1.0cm}
\begin{sideways}\begin{minipage}[b]{\textheight}
\tiny
\begin{center}
\vspace{1.5cm} 
\hspace{-1.0cm}
\begin{tabular}{|c|c|c|c|c|c|c|c|c|c|c|c|c|c|c|c|c|c|c|c|c|c|c|c|c|c|c|c|c|c|c|}
\multicolumn{1}{c}{ \begin{rotate}{45} $Q^2$ [GeV$^2$]\end{rotate}} 
& \multicolumn{1}{c}{ \begin{rotate}{45} sample \end{rotate}} 
& \multicolumn{1}{c}{ \begin{rotate}{45} $x$ \end{rotate}}
& \multicolumn{1}{c}{ \begin{rotate}{45} $F_2/\alpha$ \end{rotate}}
& \multicolumn{1}{c}{ \begin{rotate}{45} statistical error \end{rotate}}
& \multicolumn{1}{c}{ \begin{rotate}{45} total systematic + \end{rotate}}
& \multicolumn{1}{c}{ \begin{rotate}{45} total systematic - \end{rotate}}
& \multicolumn{1}{c}{ \begin{rotate}{45} MC modelling \end{rotate}} 
& \multicolumn{1}{c}{ \begin{rotate}{45} unfolding method \end{rotate}} 
& \multicolumn{1}{c}{ \begin{rotate}{45} unfolding bins \end{rotate}}
& \multicolumn{1}{c}{ \begin{rotate}{45} $W$ factor + \end{rotate}} 
& \multicolumn{1}{c}{ \begin{rotate}{45} $W$ factor - \end{rotate}} 
& \multicolumn{1}{c}{ \begin{rotate}{45} \etag,min + \end{rotate}} 
& \multicolumn{1}{c}{ \begin{rotate}{45} \etag,min - \end{rotate}} 
& \multicolumn{1}{c}{ \begin{rotate}{45} \ttag,min \end{rotate}} 
& \multicolumn{1}{c}{ \begin{rotate}{45} \ttag,max \end{rotate}} 
& \multicolumn{1}{c}{ \begin{rotate}{45} \wfd,min \end{rotate}} 
& \multicolumn{1}{c}{ \begin{rotate}{45} \wfd,max - \end{rotate}} 
& \multicolumn{1}{c}{ \begin{rotate}{45} \wfd,max + \end{rotate}} 
& \multicolumn{1}{c}{ \begin{rotate}{45} \ea,max - \end{rotate}} 
& \multicolumn{1}{c}{ \begin{rotate}{45} \ea,max + \end{rotate}} 
& \multicolumn{1}{c}{ \begin{rotate}{45} phi region \end{rotate}}
& \multicolumn{1}{c}{ \begin{rotate}{45} calibration + \end{rotate}}
& \multicolumn{1}{c}{ \begin{rotate}{45} calibration - \end{rotate}} 
& \multicolumn{1}{c}{ \begin{rotate}{45} ECAL scale + \end{rotate}}
& \multicolumn{1}{c}{ \begin{rotate}{45} ECAL scale - \end{rotate}} 
& \multicolumn{1}{c}{ \begin{rotate}{45} background + \end{rotate}}
& \multicolumn{1}{c}{ \begin{rotate}{45} background - \end{rotate}}
& \multicolumn{1}{c}{ \begin{rotate}{45} bin centre \end{rotate}} 
\\ \hline

1.9 & LEP1 SW
 & 0.0012 & 0.269 & 10 & 6.6 & 12.8
 & 0.18 & 0 & -2.01 & 4.37 & -9.08 & 0 & 0 & -4.83 & 0 & 0
 & 1.42 & 0.81 & -3.03 & 0 & 0 & -5.61 & 3.36
 & -3.26 & 2.1 & 1.33 & -1.72 & 1.54 \\
 &
 & 0.0063 & 0.177 & 5.1 & 9.6 & 8
 & -7.11 & 1.73 & 0 & -1.51 & 1.01 & -2.78 & 5.94 & 0 & 0 & 0
 & -0.19 & -0.17 & -1.54 & 0 & 0 & 0.03 & -0.19
 & -1.08 & 1.28 & -0.03 & 0.01 & 0.05 \\
 &
 & 0.0321 & 0.179 & 3.8 & 4.1 & 3.6
 & -3.49 & 0 & 0 & 1.19 & 0 & 0 & 0.12 & 0 & 0 & 0
 & 0.05 & 0.04 & 0.3 & 0.05 & 0 & -0.25 & 0.65
 & -0.78 & 1.66 & 0.04 & -0.03 & 0.02 \\
 &
 & 0.1124 & 0.227 & 4.2 & 5.4 & 5.3
 & 3.39 & -2.36 & 0 & -2.06 & 0 & 0.48 & -0.56 & 0 & 0 & 3.62
 & -0.02 & -0.01 & -0.19 & -0.04 & 0 & 1.8 & -1.71
 & 0.65 & -1.73 & -0.02 & 0.02 & 0.45 \\
\hline 
 3.7 & LEP1 SW
 & 0.0032 & 0.269 & 12.2 & 17.3 & 12.4
 & -4.23 & 0 & 11.86 & 8.31 & -1.62 & 0 & 0 & 0 & 0 & 0
 & -9.83 & -1.42 & 4.7 & 0 & 0 & -3.18 & 4.06
 & -3.51 & 4.53 & 2.77 & -3.55 & 1.36 \\
 &
 & 0.0143 & 0.232 & 5.6 & 10 & 9.2
 & -8.29 & 4.28 & 0 & 0 & -0.61 & 0 & 0 & 0 & -2.43 & 0.98
 & 1.58 & 0.22 & -1.47 & 0.1 & 0 & -2.18 & 2.69
 & -1.41 & 1.22 & 0.12 & -0.27 & 0.35 \\
 &
 & 0.0639 & 0.259 & 3.9 & 2.4 & 4.9
 & -0.5 & -0.84 & -3.35 & 0 & 0 & 0 & -0.92 & 0 & 0.15 & 0
 & -0.36 & -0.07 & 0.6 & -0.16 & 0 & -2.35 & 1.94
 & -2.37 & 1.25 & -0.01 & 0.04 & 0.1 \\
 &
 & 0.1986 & 0.296 & 4.6 & 9.9 & 7.3
 & 6.59 & -1.68 & 6.01 & -1.01 & 0 & 0 & 0.32 & 0 & 0 & 0
 & 0.17 & 0.01 & -0.26 & 0 & 0 & -1.67 & 2.21
 & 3.71 & -1.71 & -0.01 & -0.02 & 0.54 \\
\hline
 8.9 & LEP1 FD
 & 0.0235 & 0.221 & 7.6 & 13.5 & 11.6
 & -10.51 & 0 & 3.85 & 6.41 & 1.95 & -1.65 & 0 & 0 & 0 & -0.96
 & 0 & 0.01 & -1.02 & 0 & 0 & -2.12 & 1.29
 & -2.68 & 2.31 & 2.14 & -2.76 & 0.84 \\
 &
 & 0.1054 & 0.308 & 4.6 & 3.6 & 3.9
 & 0.64 & -2.68 & 0 & 0 & -2.05 & 0 & 0 & 2.47 & 0 & 0.53
 & 0 & 0 & 0 & 0 & 0 & -0.42 & 0.62
 & -1.75 & 2.29 & -0.36 & 0.46 & 0.06 \\
 &
 & 0.3331 & 0.379 & 5.8 & 4.5 & 4
 & -2.63 & 1.67 & 0 & 0 & 1.04 & 0 & 0 & 0 & 0 & 0
 & 0 & 0 & 0 & 0.08 & 0 & -0.24 & -0.74
 & 3.04 & -2.91 & 0.13 & -0.22 & 0.37 \\
\hline 
 10.7 & LEP2 SW
 & 0.0021 & 0.362 & 12.3 & 16.1 & 10.7
 & -1.21 & 0 & 0 & 4.23 & -3.94 & -5.05 & 9.62 & 3.5 & 0 & 0
 & 0 & 3.45 & 0 & 4.08 & 7.41 & -8.11 & 6.85
 & -2.34 & 0.27 & 0 & 0 & 1.54 \\
 &
 & 0.0117 & 0.263 & 5.8 & 12.3 & 11.6
 & -9.85 & 5.28 & 3.68 & 0 & 0 & 0 & 0 & -5.96 & 0 & 0.35
 & 0 & -0.4 & 0 & 0 & 3.27 & 0.67 & -0.22
 & -0.86 & 1.3 & 0 & 0 & 0.5 \\
 &
 & 0.0639 & 0.275 & 4 & 10.4 & 10.8
 & -9.14 & 0 & -4.5 & 0 & 0 & -2.17 & 0.64 & 4.44 & 0 & -2.32
 & 0 & 0.09 & 0 & 0 & 1.14 & 0.83 & -0.45
 & -1.5 & 1.64 & 0 & 0 & 0.41 \\
 &
 & 0.3143 & 0.351 & 3.5 & 7 & 4.5
 & -3.68 & 0 & 0 & 0 & 0 & 0 & 0 & 2.74 & 0 & 3.87
 & 0 & -0.05 & 0.02 & 0 & 2.57 & 1.74 & -1.83
 & 1.76 & -1.87 & 0 & 0 & 0.17 \\
\hline 
 17.5 & LEP1 FD
 & 0.0439 & 0.273 & 10.4 & 11.8 & 14.4
 & -8.93 & 0 & 0 & 0 & -1.21 & -3.38 & 0 & 0 & 2.55 & 0.95
 & -0.32 & -0.03 & 0 & -6.56 & 0 & -3.43 & 2.75
 & -2.29 & 3.87 & 5.23 & -7.45 & 0.96 \\
 &
 & 0.1534 & 0.375 & 6.1 & 5.4 & 3.6
 & -1.34 & 0 & 3.25 & 0 & 0 & 0.52 & 1.26 & 0 & 0.66 & 0
 & 0.23 & -0.02 & -0.92 & 1.25 & 0 & -2.48 & 3.06
 & -1.99 & 1.75 & -0.14 & -0.27 & 0.03 \\
 &
 & 0.3945 & 0.501 & 5.4 & 5.3 & 3.8
 & -0.78 & -1.33 & -1.54 & 0 & 0 & 0 & 0 & 0 & 0 & 3.71
 & -0.11 & -0.02 & 0 & 0 & 0 & -2.72 & 2.76
 & 1.53 & -1.61 & -0.05 & 0.25 & 1.42 \\
\hline 
 17.8 & LEP2 SW
 & 0.0033 & 0.428 & 14.2 & 12.8 & 16.7
 & -2.22 & -2.88 & -2.91 & 0 & 0 & -15.09 & 2.33 & 0 & 0 & -0.57
 & 0 & 0 & 2.89 & 0 & 0.68 & -5.28 & 11.84
 & -1.03 & 1.26 & 0 & 0 & 1.33 \\
 &
 & 0.0166 & 0.295 & 6.4 & 11.2 & 6.8
 & -5.79 & 6.79 & 0 & 2.84 & 0 & 2.04 & 1.95 & 0 & -0.78 & 1.03
 & 0 & 0.52 & 0 & 0 & 4.49 & -3.41 & 2.63
 & -0.99 & 0.99 & 0 & 0 & 0.81 \\
 &
 & 0.0821 & 0.336 & 3.9 & 12.3 & 12.6
 & -11.56 & -1.87 & 0 & -0.19 & 0 & -1.68 & 0 & 0 & 0 & -2.66
 & 0 & -0.08 & -0.48 & 0.43 & -1.05 & -2.41 & 3.32
 & -2.25 & 2.46 & 0 & 0 & 0.52 \\
 &
 & 0.3483 & 0.43 & 2.9 & 7.5 & 5.9
 & -3.7 & 0 & 0 & 0 & 0 & -0.19 & 0.18 & 0 & -3.29 & 5.65
 & 0 & 0.02 & 0.18 & -0.12 & 0.85 & -2.47 & 2.46
 & 1.86 & -2.04 & 0 & 0 & 0.7 \\
 \hline

\end{tabular}
\label{tab:errxq}
\caption{Systematic errors for \ftn\ as a function of $x$ 
         in bins of \qsq. The statistical error and the contributions 
         to the systematic error from each of the sources are listed as
         percentages of \ftn. The systematic errors are corrected for
         the expected statistical component as described in the
         text. They are summed in quadrature separately for 
         positive and negative
         deviations, except for `MC modelling' and `bin centre' 
         which are symmetric.}
\end{center}
\end{minipage}\end{sideways}
\end{table}




\begin{figure}[p]
\vspace{-8cm}
\begin{center}
\epsfig{bbllx=50,bblly=0,bburx=595,bbury=842,file=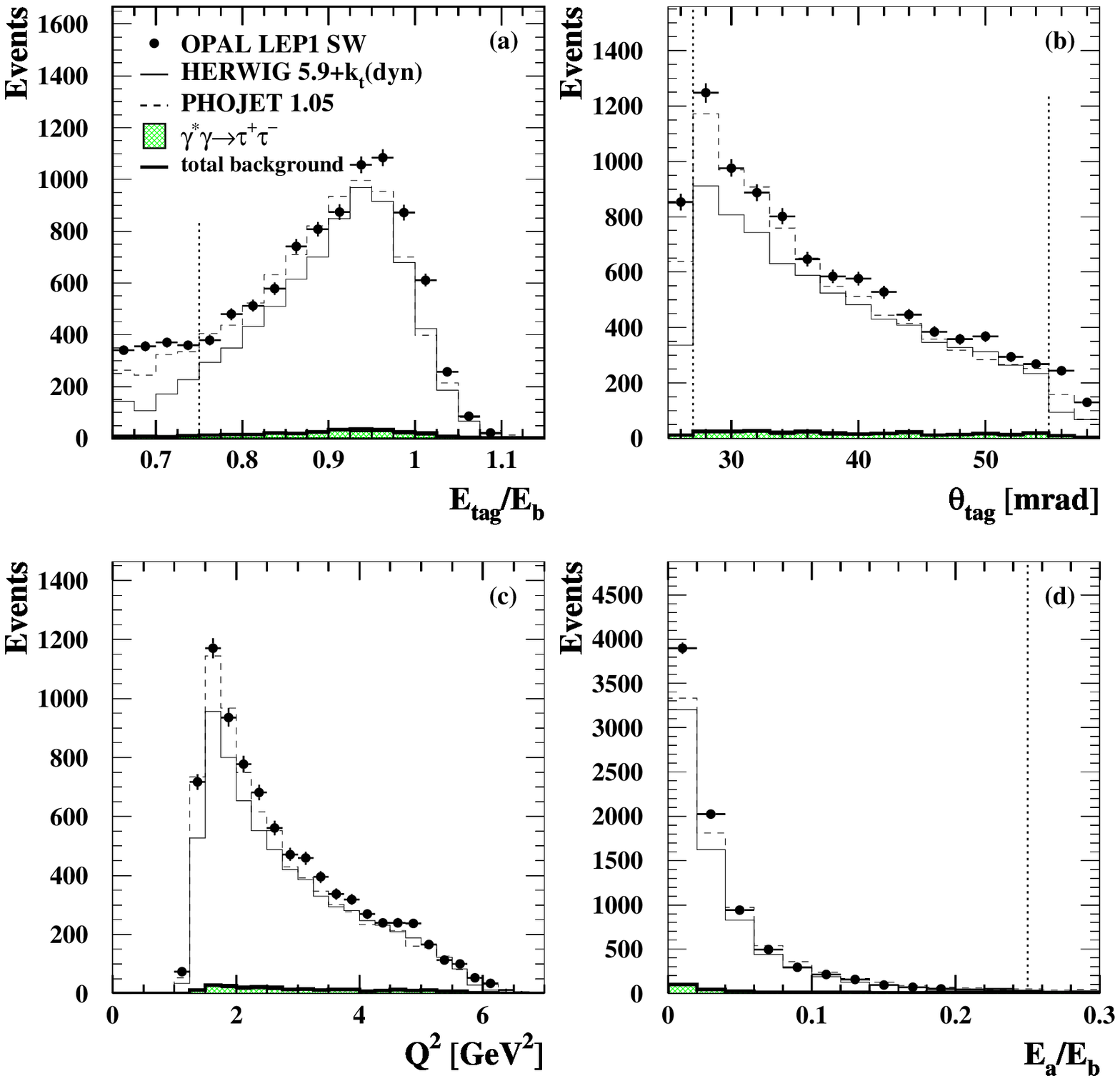,height=28cm}
\vspace{-4.5cm}
\caption{Comparison of data distributions with Monte Carlo predictions for the
         LEP1 SW sample. 
         The dominant background source,
         \ggtau, the total background
         and the sum of the signal and the total background for 
         HERWIG 5.9+\kt\ and PHOJET 1.05 are shown.
         The Monte Carlo samples have
         been normalised to the data luminosity.
         All selection cuts have been applied, except
         for any cut on the variable in the plot. The
         cuts are shown as dotted lines. 
         The distributions shown are:
         a) \etag/\eb, the energy of the tagged electron as a fraction
            of the beam energy, 
         b) \ttag, the polar angle of the tagged electron,
         c) the measured \qsq\, and
         d) \ea/\eb, the energy of the most energetic
            electromagnetic cluster in
            the hemisphere opposite the tagged electron, as
            a fraction of the beam energy.
         The errors are statistical only.
         }\label{91swgen1}
\end{center}
\end{figure}


\begin{figure}[p]
\vspace{-8cm}
\begin{center}
\epsfig{bbllx=50,bblly=0,bburx=595,bbury=842,file=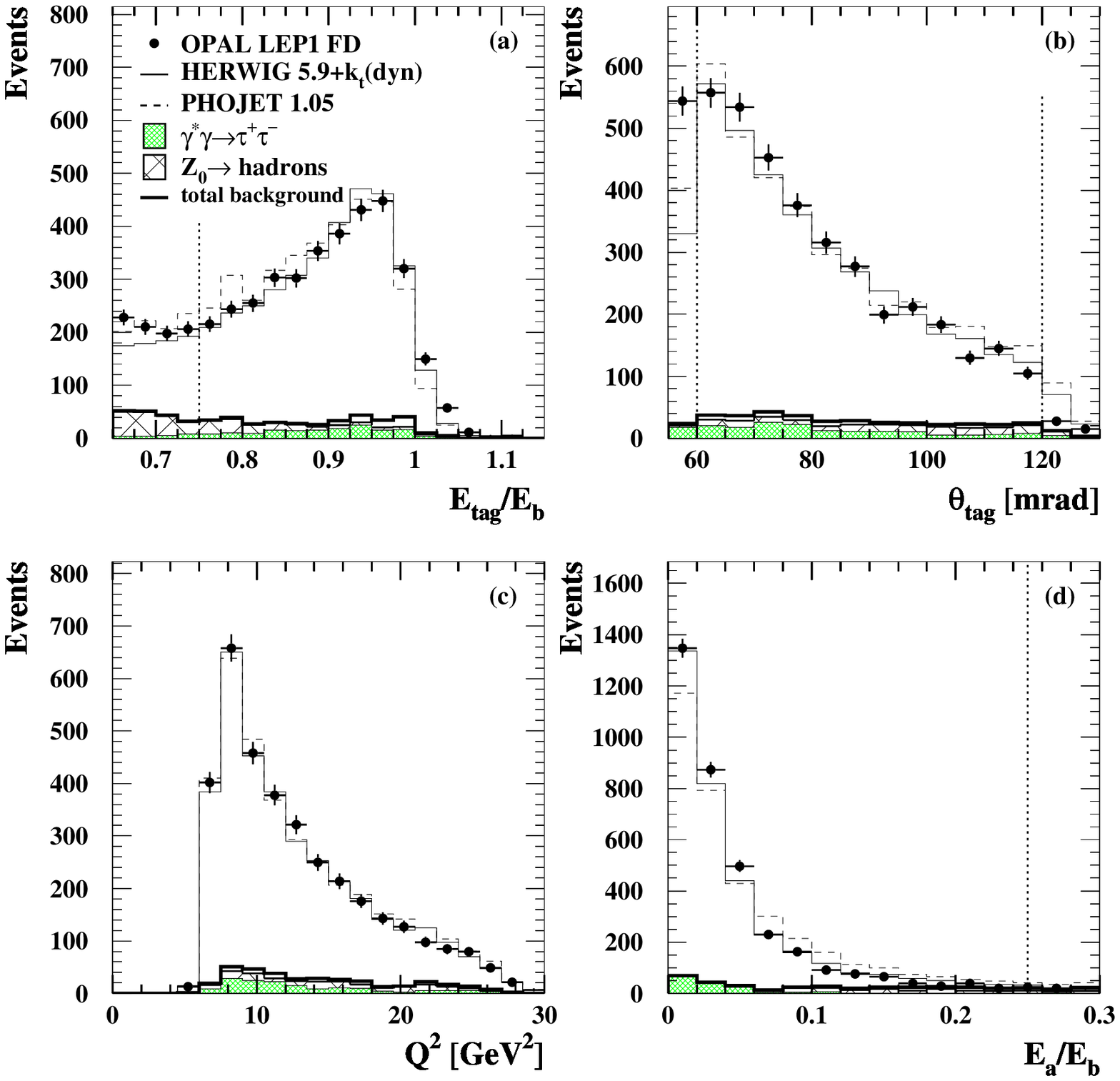,height=28cm}
\vspace{-4.5cm}
\caption{Comparison of data distributions with Monte Carlo predictions for the
         LEP1 FD sample. 
         The dominant background sources,
         \ggtau\ and \znhad, the total background
         and the sum of the signal and the total background for 
         HERWIG 5.9+\kt\ and PHOJET 1.05 are shown.
         The Monte Carlo samples have
         been normalised to the data luminosity.
         All selection cuts have been applied, except
         for any cut on the variable in the plot. The
         cuts are shown as dotted lines.
         The variables in the plots are as defined in
         Figure~\ref{91swgen1}.
         The errors are statistical only.
        }\label{91fdgen1}
\end{center}
\end{figure}


\begin{figure}[p]
\vspace{-8cm}
\begin{center}
\epsfig{bbllx=50,bblly=0,bburx=595,bbury=842,file=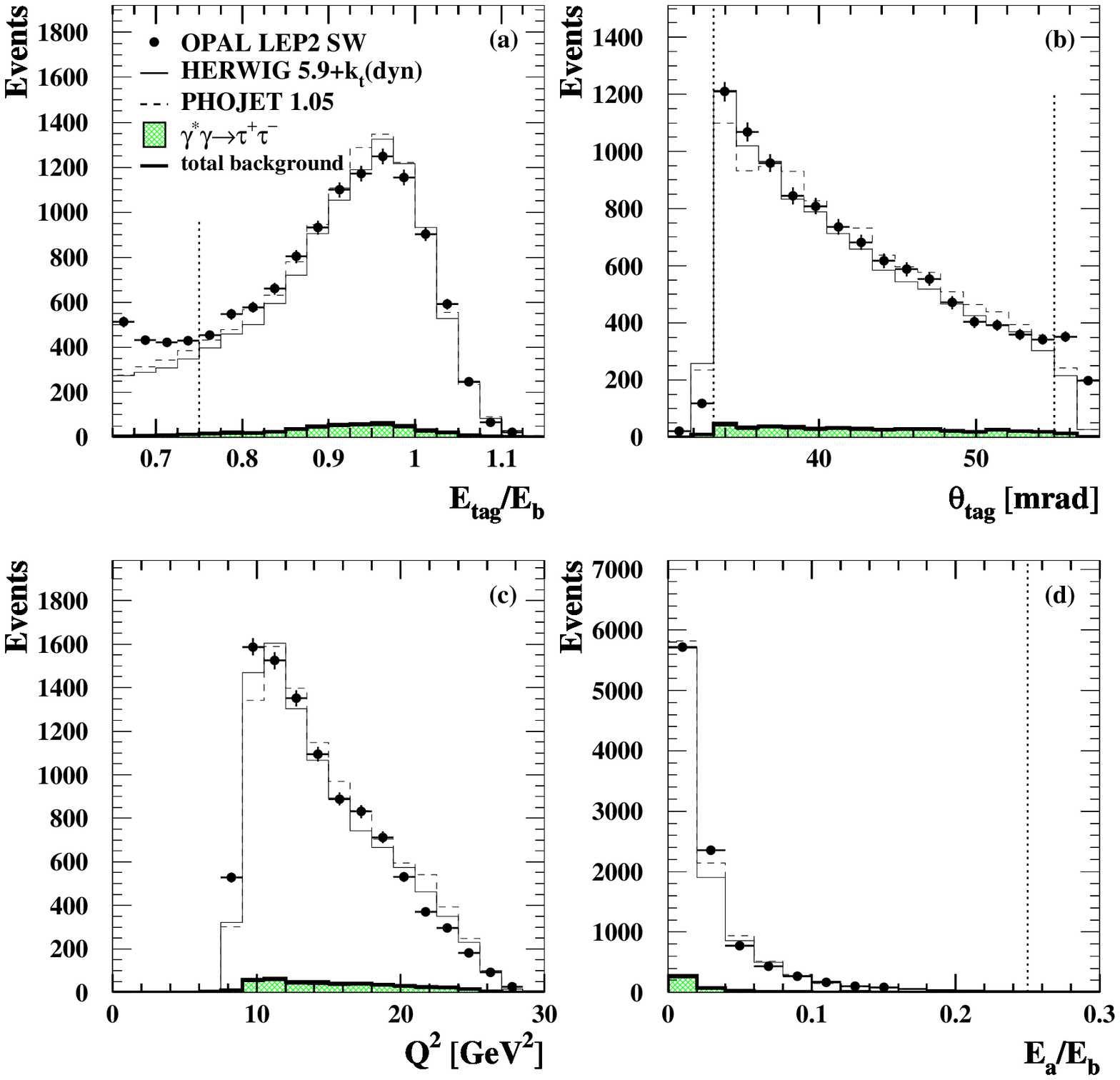,height=28cm}
\vspace{-4.5cm}
\caption{Comparison of data distributions with Monte Carlo predictions
         for the LEP2 SW sample. 
         The dominant background source,
         \ggtau, the total background
         and the sum of the signal and the total background for 
         HERWIG 5.9+\kt\ and PHOJET 1.05 are shown.
         The Monte Carlo samples have
         been normalised to the data luminosity.
         All selection cuts have been applied, except
         for any cut on the variable in the plot. The
         cuts are shown as dotted lines.
         The variables in the plots are as defined in Figure~\ref{91swgen1}.
         The errors are statistical only.
        }\label{189swgen1}
\end{center}
\end{figure}


\begin{figure}[p]
\vspace{-8cm}
\begin{center}
\epsfig{bbllx=50,bblly=0,bburx=595,bbury=842,file=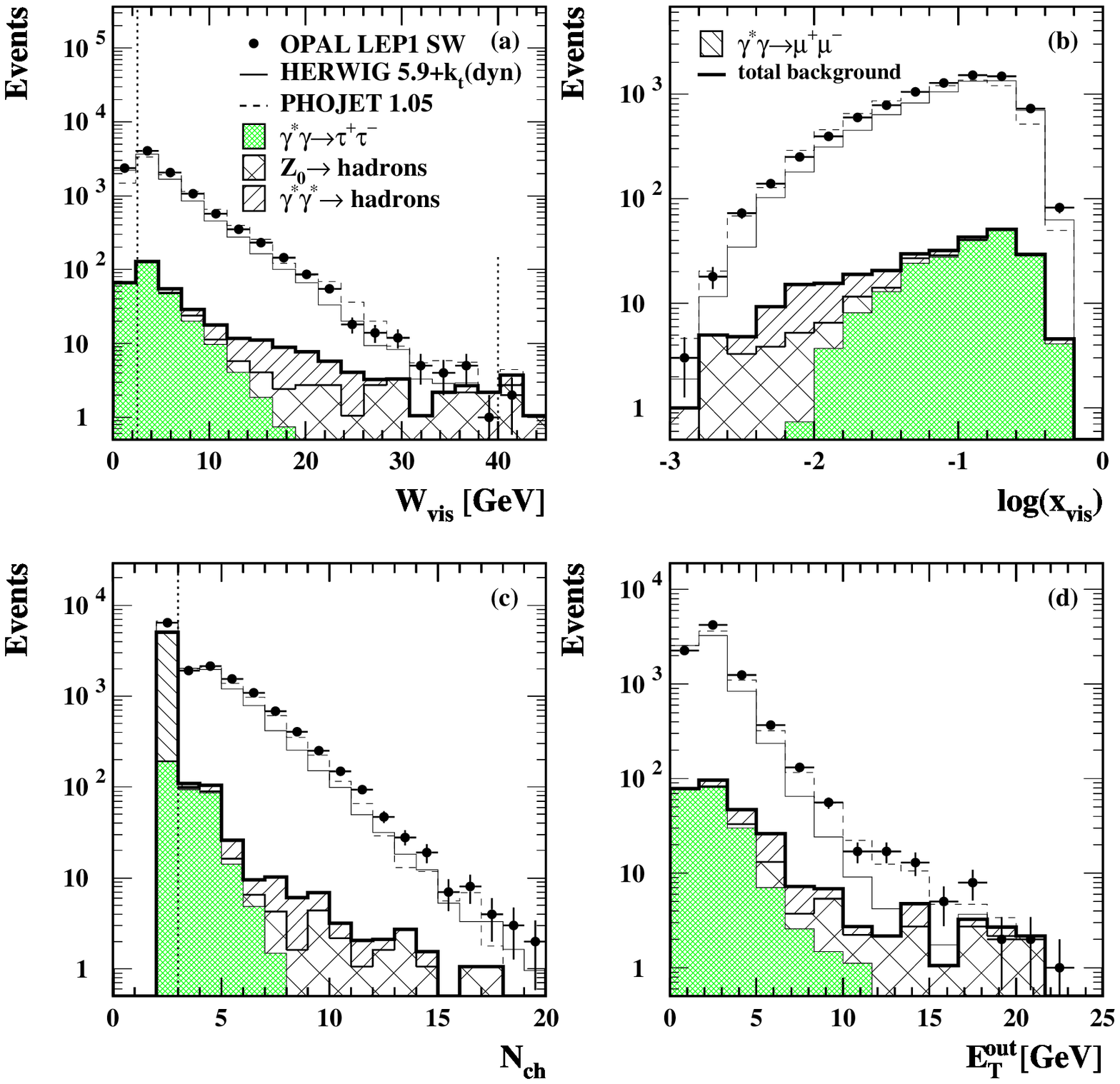,height=28cm}
\vspace{-4.5cm}
\caption{Comparison of data distributions with Monte Carlo predictions
         for the LEP1 SW sample. 
         The dominant background sources,
         \ggtau, \znhad\ and \gsgshad\ (with \psq$>$1.0 \gevsq), the total background
         and the sum of the signal and the total background for 
         HERWIG 5.9+\kt\ and PHOJET 1.05 are shown.
         For illustration, the background from \ggmu\ is also shown,
         though it is only significant below the \nch\ cut.
         The Monte Carlo samples have
         been normalised to the data luminosity.
         All selection cuts have been applied, except
         for any cut on the variable in the plot.
         The cuts are shown as dotted lines. 
         The distributions shown are:
         a) \wfd, the measured invariant mass of the hadronic final
         state,
         b) $\log{(\xfd)}$, the logarithm of the measured value of $x$.
         c) \nch, the number of tracks in the event and
         d) \etout, the transverse component of energy out of the tag plane.
         The errors are statistical only.
        }\label{91swgen2}
\end{center}
\end{figure}


\begin{figure}[p]
\vspace{-8cm}
\begin{center}
\epsfig{bbllx=50,bblly=0,bburx=595,bbury=842,file=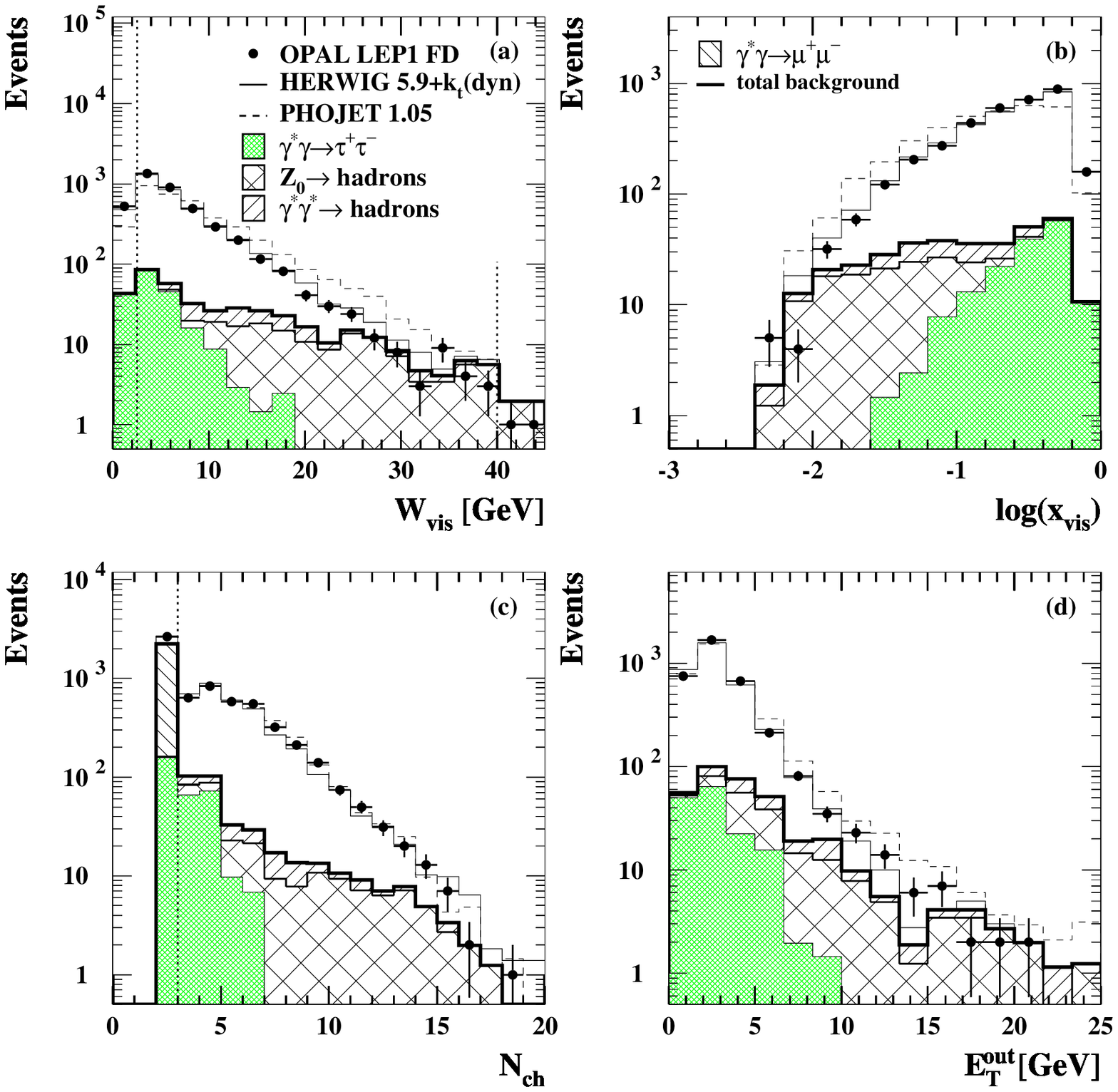,height=28cm}
\vspace{-4.5cm}
\caption{Comparison of data distributions with Monte Carlo predictions
         for the LEP1 FD sample. 
         The dominant background sources,
         \ggtau, \znhad\ and \gsgshad\ (with \psq$>$1.0 \gevsq), the total background
         and the sum of the signal and the total background for 
         HERWIG 5.9+\kt\ and PHOJET 1.05 are shown.
         For illustration, the background from \ggmu\ is also shown,
         though it is only significant below the \nch\ cut.
         The Monte Carlo samples have
         been normalised to the data luminosity.
         All selection cuts have been applied, except
         for any cut on the variable in the plot.
         The cuts are shown as dotted lines. 
         The variables in the plots are as defined in Figure~\ref{91swgen2}.
         The errors are statistical only.
        }\label{91fdgen2}
\end{center}
\end{figure}


\begin{figure}[p]
\vspace{-8cm}
\begin{center}
\epsfig{bbllx=50,bblly=0,bburx=595,bbury=842,file=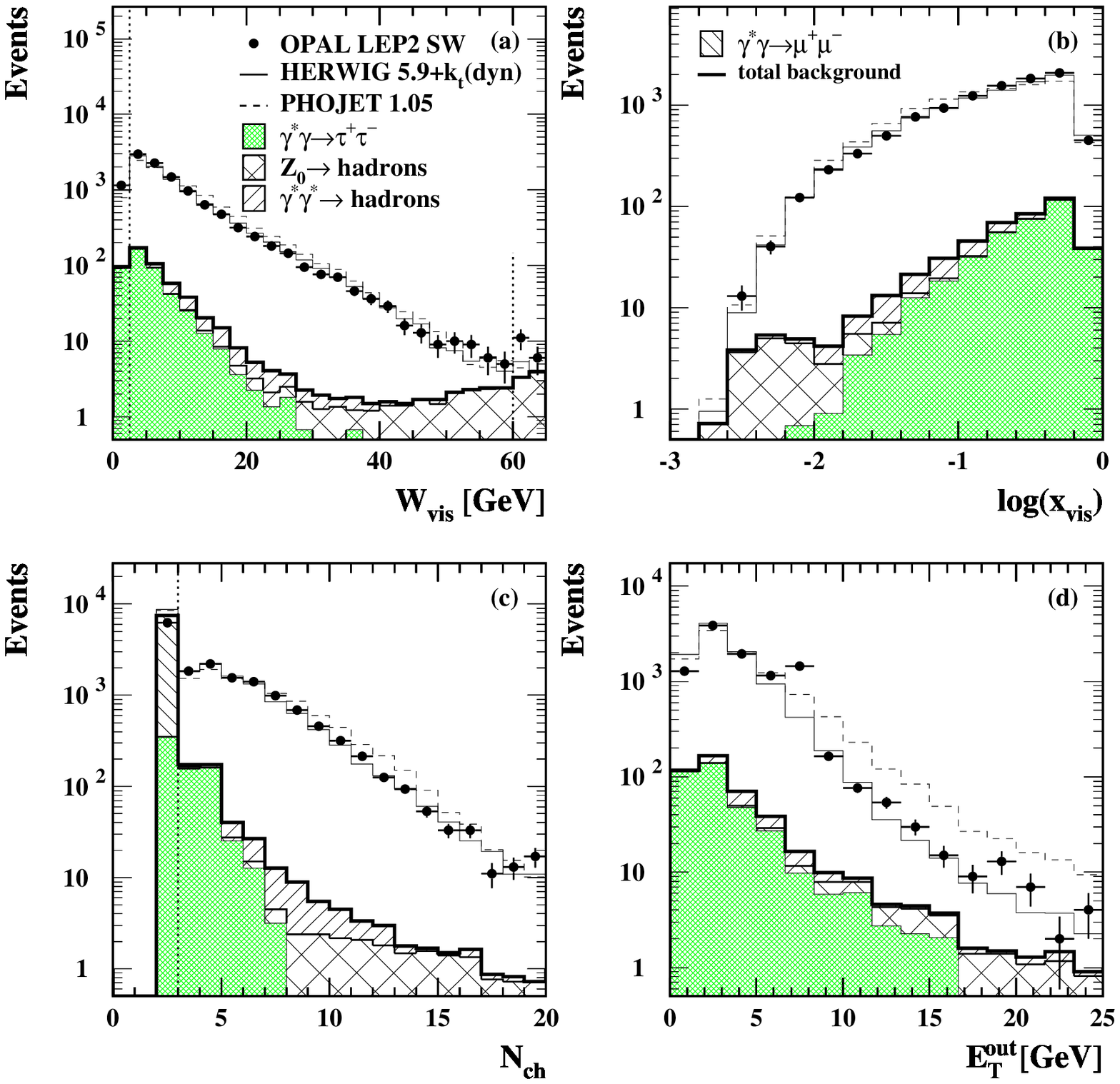,height=28cm}
\vspace{-4.5cm}
\caption{Comparison of data distributions with Monte Carlo predictions
         for the LEP2 SW sample. 
         The dominant background sources,
         \ggtau, \znhad\ and \gsgshad\ (with \psq$>$4.5 \gevsq), the total background
         and the sum of the signal and the total background for 
         HERWIG 5.9+\kt\ and PHOJET 1.05 are shown.
         For illustration, the background from \ggmu\ is also shown,
         though it is only significant below the \nch\ cut.
         The Monte Carlo samples have
         been normalised to the data luminosity.
         All selection cuts have been applied, except
         for any cut on the variable in the plot.
         The cuts are shown as dotted lines. 
         The variables in the plots are as defined in Figure~\ref{91swgen2}.
         The errors are statistical only.
        }\label{189swgen2}
\end{center}
\end{figure}


\begin{figure}[p]
\vspace{-8cm}
\begin{center}
\epsfig{bbllx=50,bblly=0,bburx=595,bbury=842,file=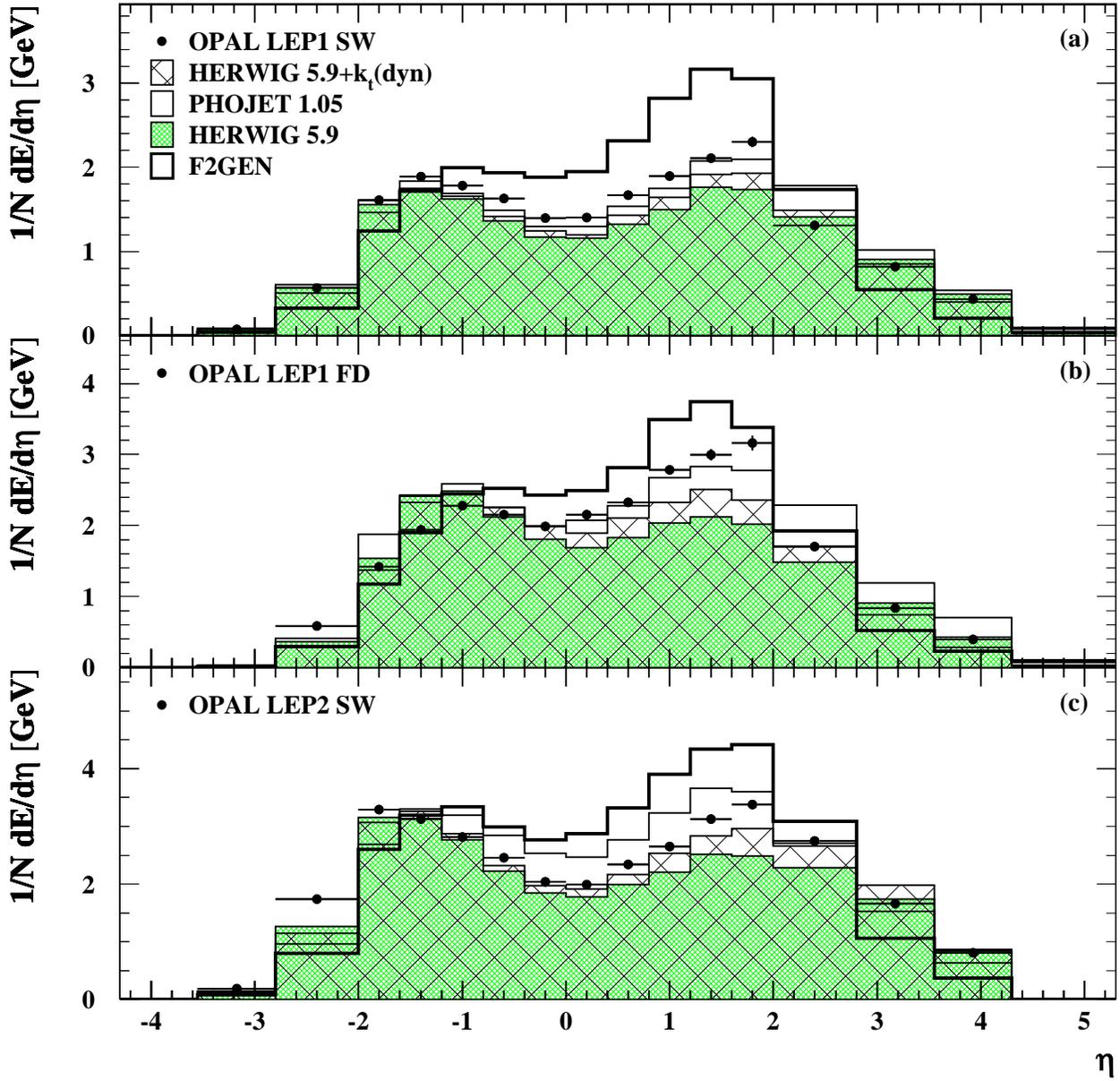,height=28cm}
\vspace{-4.5cm}
\caption{Comparison of hadronic energy flow per event for data and Monte Carlo 
         as a function of the pseudorapidity, $\eta=-\log(\tan(\theta/2))$, 
         where the polar angle
         $\theta$ is measured with respect to the beam axis on the
         side of the tagged electron.
         All three samples, LEP1 SW, LEP1 FD and LEP2 SW
         are shown.
         The tagged electron is not included in these plots.
         The errors are statistical only.
        }\label{eflows}
\end{center}
\end{figure}


\begin{figure}[p]
\vspace{-8cm}
\begin{center}
\epsfig{bbllx=50,bblly=0,bburx=595,bbury=842,file=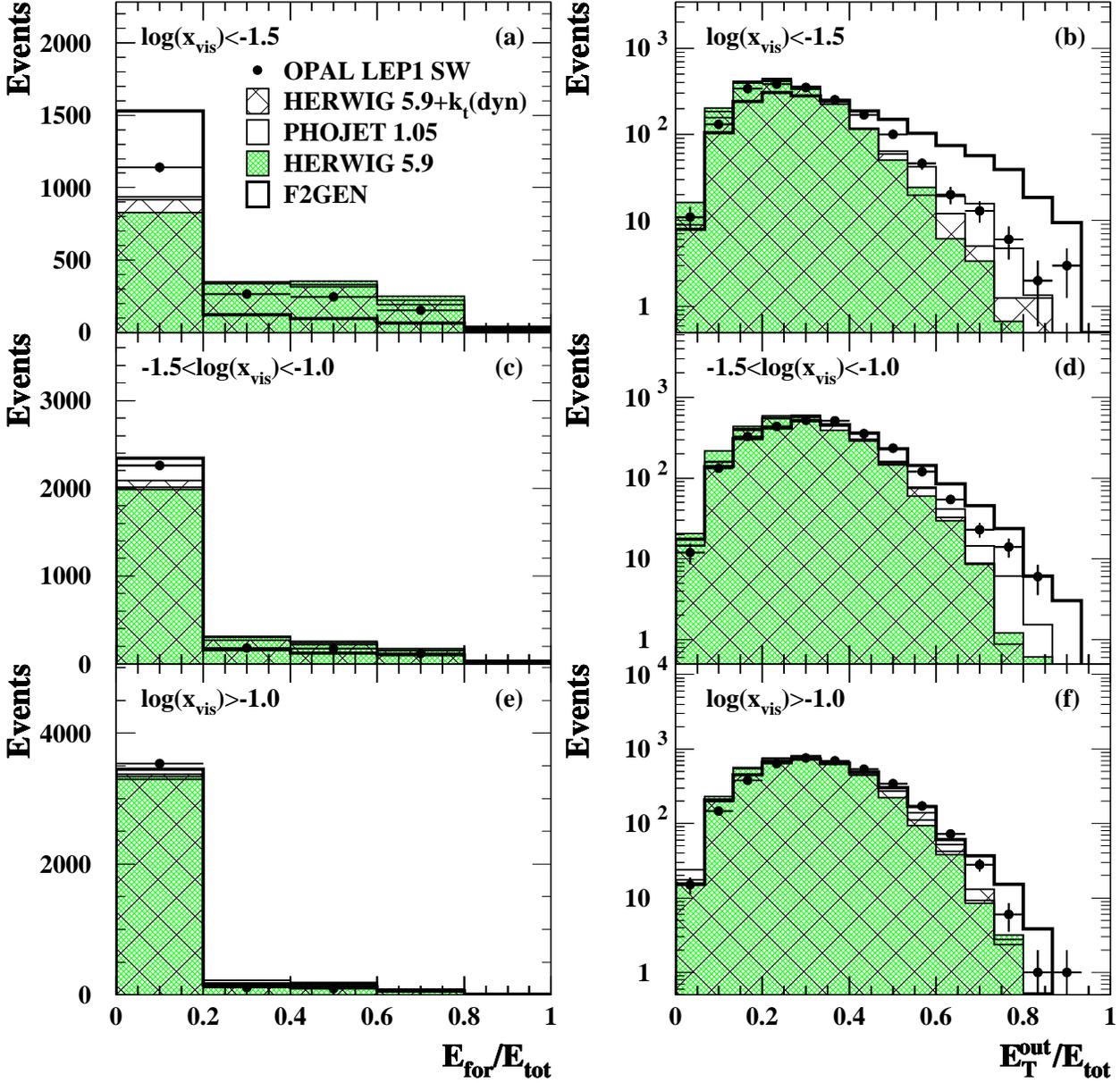,height=28cm}
\vspace{-4.5cm}
\caption{Data and signal Monte Carlo distributions for 
         two alternative variables for 
         two dimensional unfolding, for the LEP1 SW sample
         divided into three bins of \xfd.
         Plots a) c) and e) show \edist, 
         the observed energy in the forward region 
         divided by the total observed energy.
         Plots b) d) and f) show \etdist, 
         defined as the transverse component of hadronic
         energy out of the plane of the tagged electron,
         divided by the total energy. 
         The Monte Carlo samples have
         been normalised to the number of data events in each plot.
         The errors are statistical only.
        }\label{91sw2dvar}
\end{center}
\end{figure}


\begin{figure}[p]
\vspace{-8cm}
\begin{center}
\epsfig{bbllx=50,bblly=0,bburx=595,bbury=842,file=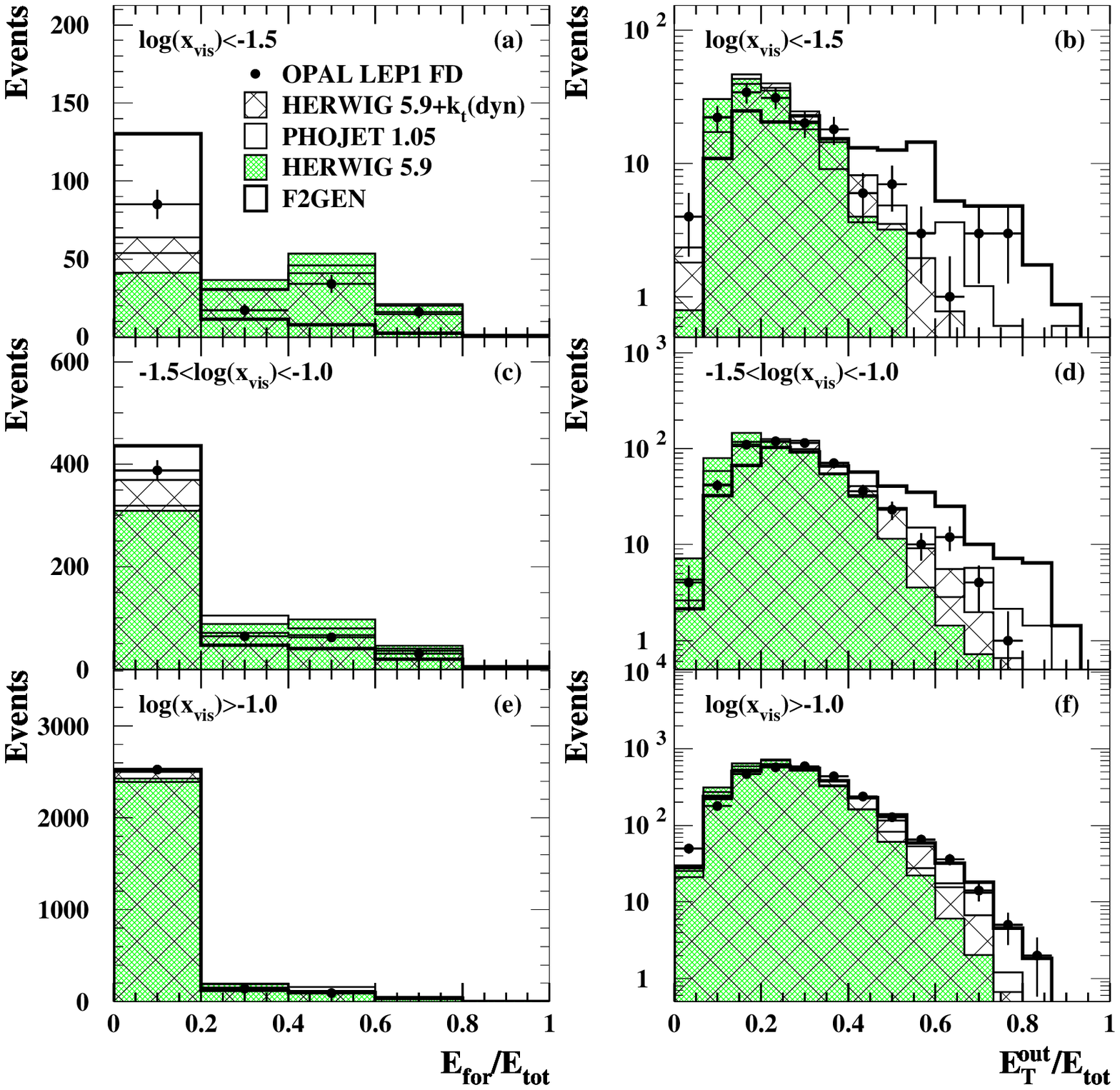,height=28cm}
\vspace{-4.5cm}
\caption{Data and signal Monte Carlo distributions for
         two alternative variables for 
         two dimensional unfolding, for the LEP1 FD sample
         divided into three bins of \xfd. 
         The variables are as defined in Figure~\ref{91sw2dvar}.
         The Monte Carlo samples have
         been normalised to the number of data events in each plot.
         The errors are statistical only.
        }\label{91fd2dvar}
\end{center}
\end{figure}


\begin{figure}[p]
\vspace{-8cm}
\begin{center}
\epsfig{bbllx=50,bblly=0,bburx=595,bbury=842,file=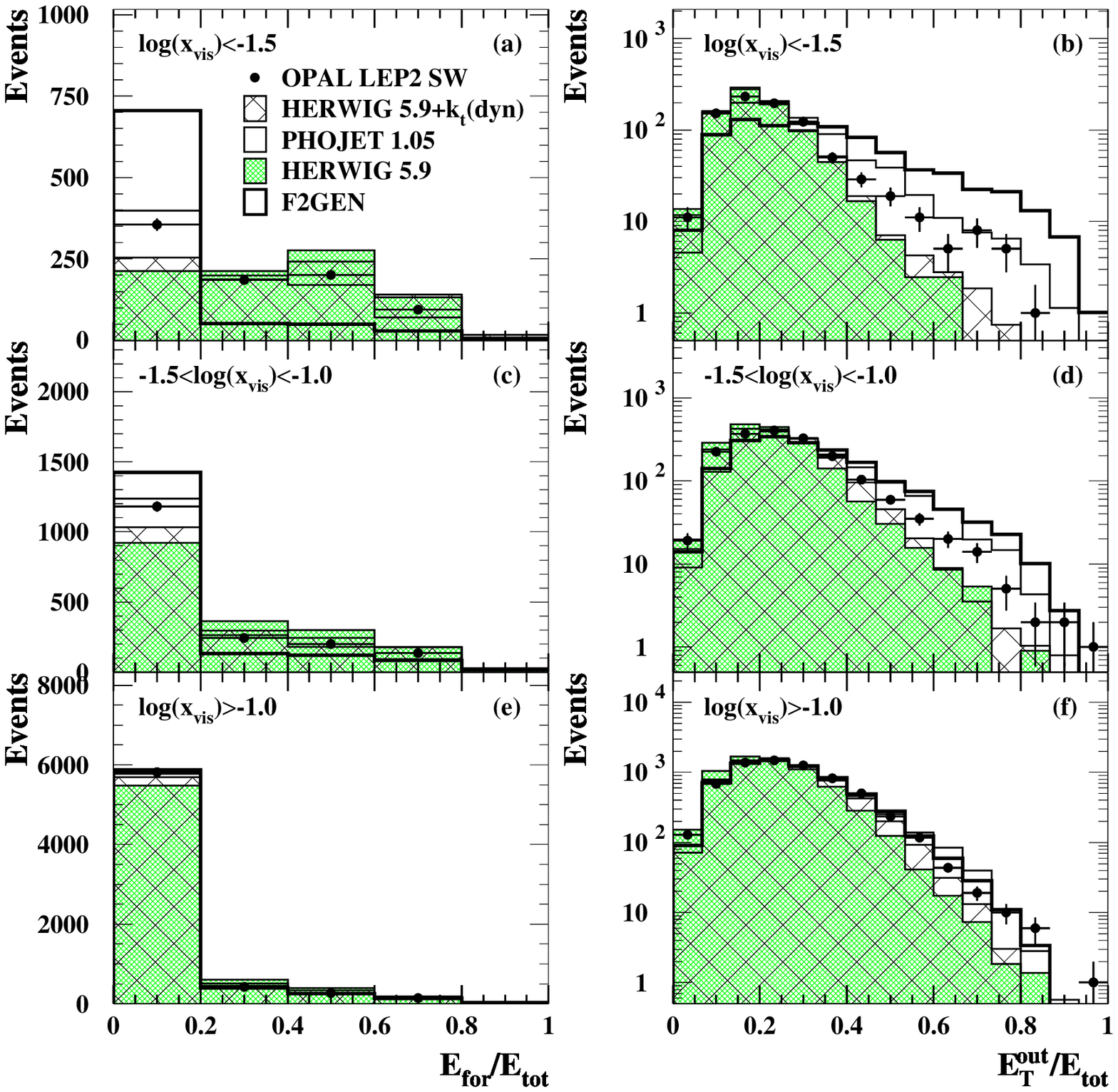,height=28cm}
\vspace{-4.5cm}
\caption{Data and signal Monte Carlo distributions for
         two alternative variables for 
         two dimensional unfolding, for the LEP2 SW sample
         divided into three bins of \xfd. 
         The variables are as defined in Figure~\ref{91sw2dvar}.
         The Monte Carlo samples have
         been normalised to the number of data events in each plot.
         The errors are statistical only.
        }\label{189sw2dvar}
\end{center}
\end{figure}


\begin{figure}[p]
\vspace{-8cm}
\begin{center}
\epsfig{bbllx=50,bblly=0,bburx=595,bbury=842,file=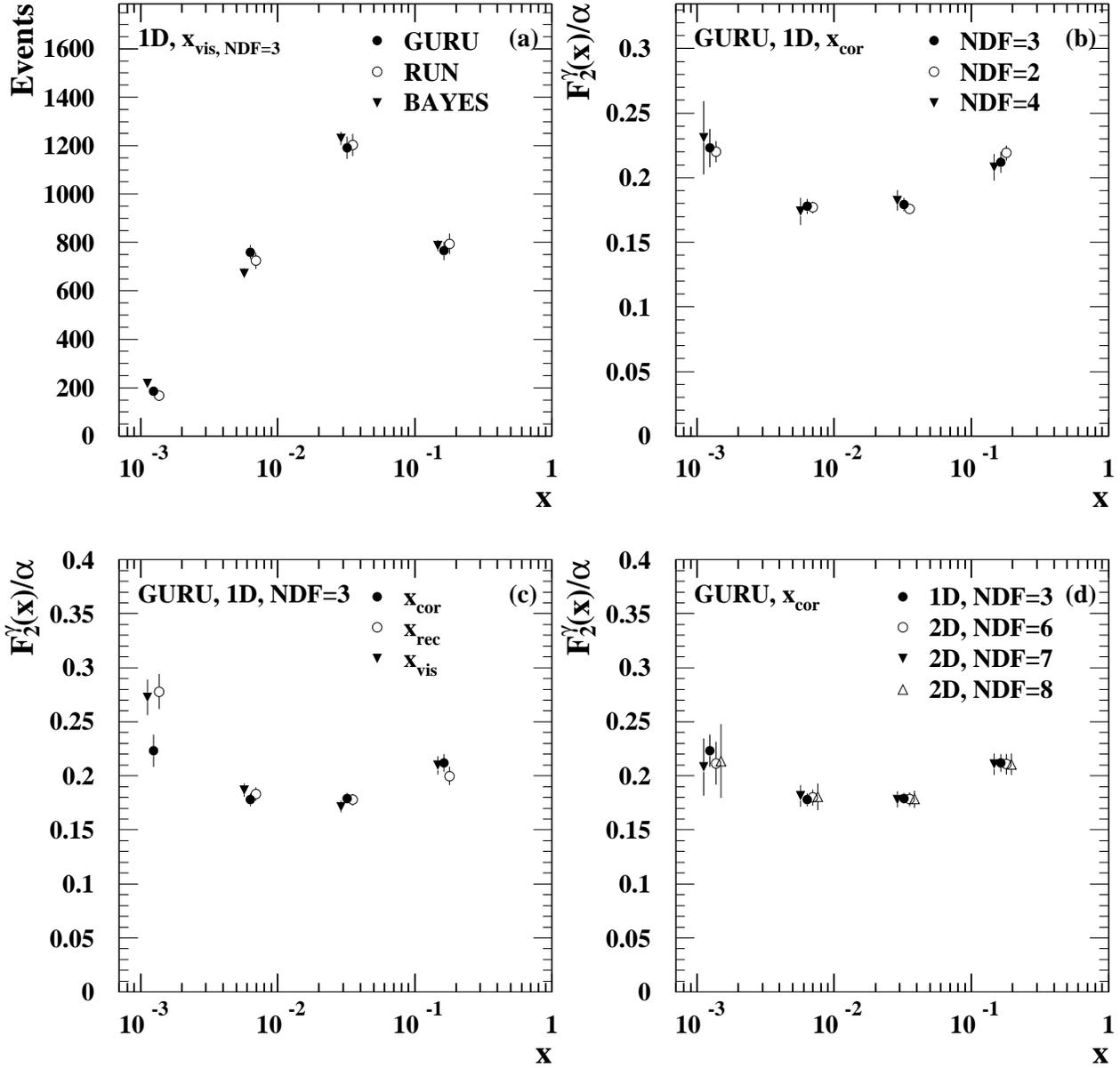,height=28cm}
\vspace{-4.5cm}
\caption{Tests of the unfolding procedure using the LEP1 SW low \qsq\
         data unfolded with the HERWIG 5.9+\kt\ Monte Carlo sample. 
         (a) Comparison of the results using GURU, RUN and BAYES.
         (b) Different degrees of freedom for one dimensional unfolding.
         (c) Different measured $x$ variables for one dimensional
             unfolding.
         (d) Different degrees of freedom for two dimensional unfolding
             using a random number as the second variable.
         The errors are statistical only.
        }\label{uf1}
\end{center}
\end{figure}


\begin{figure}[p]
\vspace{-8cm}
\begin{center}
\epsfig{bbllx=50,bblly=0,bburx=595,bbury=842,file=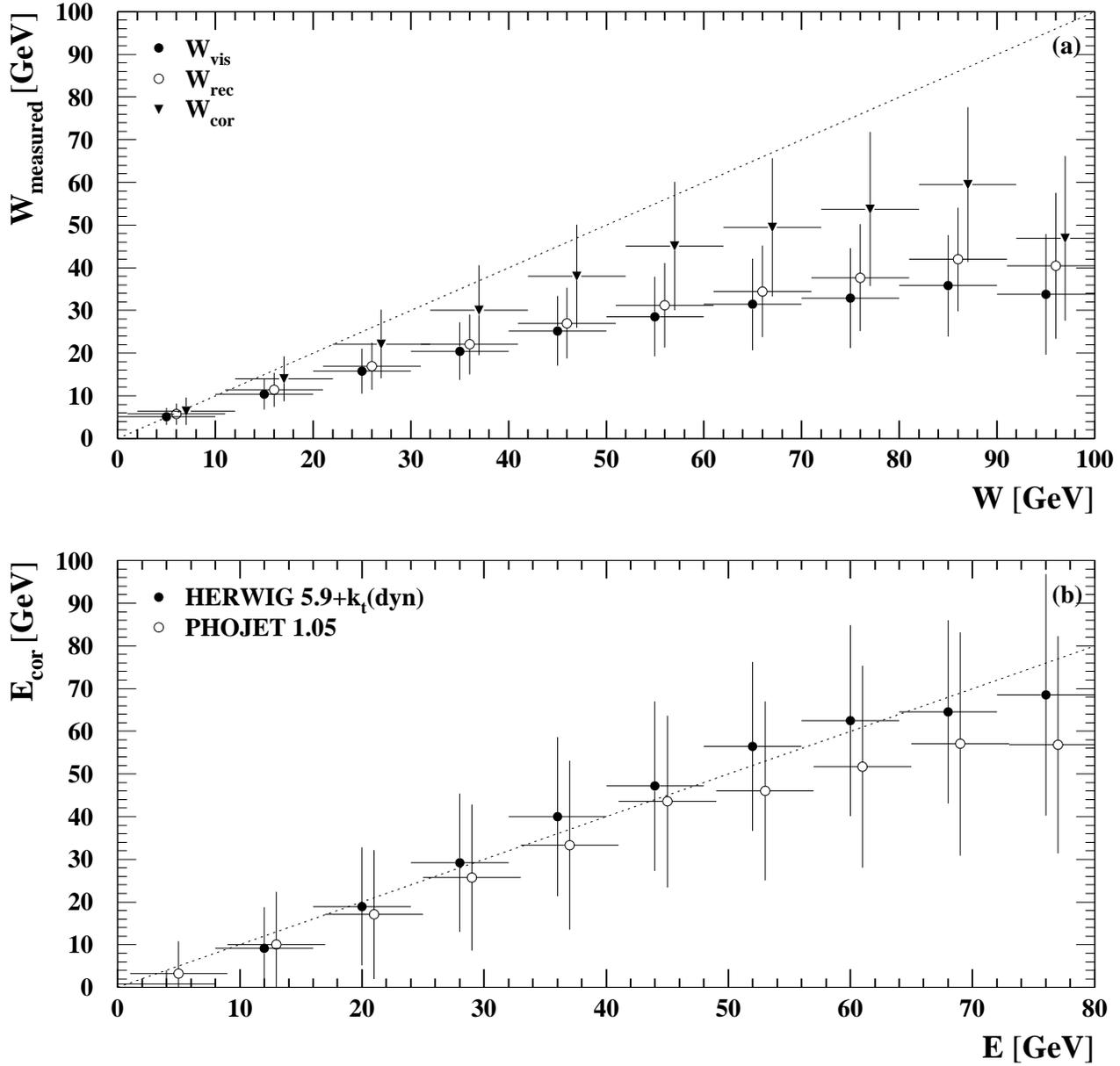,height=28cm}
\vspace{-4.5cm}
\caption{(a) The correlation between the 
         generated and measured 
         invariant mass, $W$ and $W_{\rm measured}$, for the HERWIG 5.9+\kt\ LEP2 SW 
         Monte Carlo sample,
         using, $W_{\rm vis}$, $W_{\rm rec}$ and $W_{\rm cor}$.
         (b) The corrected energy, $E_{\rm cor}$,
         (2.5 times the visible energy) in the forward region 
         against the total generated energy, $E$, deposited in that region,
         for HERWIG 5.9+\kt\ and PHOJET 1.05.
         The vertical error bars represent the RMS spread within each bin.
         The dotted lines represent perfect correlation.
         }\label{wcor}
\end{center}
\end{figure}


\begin{figure}[p]
\vspace{-8cm}
\begin{center}
\epsfig{bbllx=50,bblly=0,bburx=595,bbury=842,file=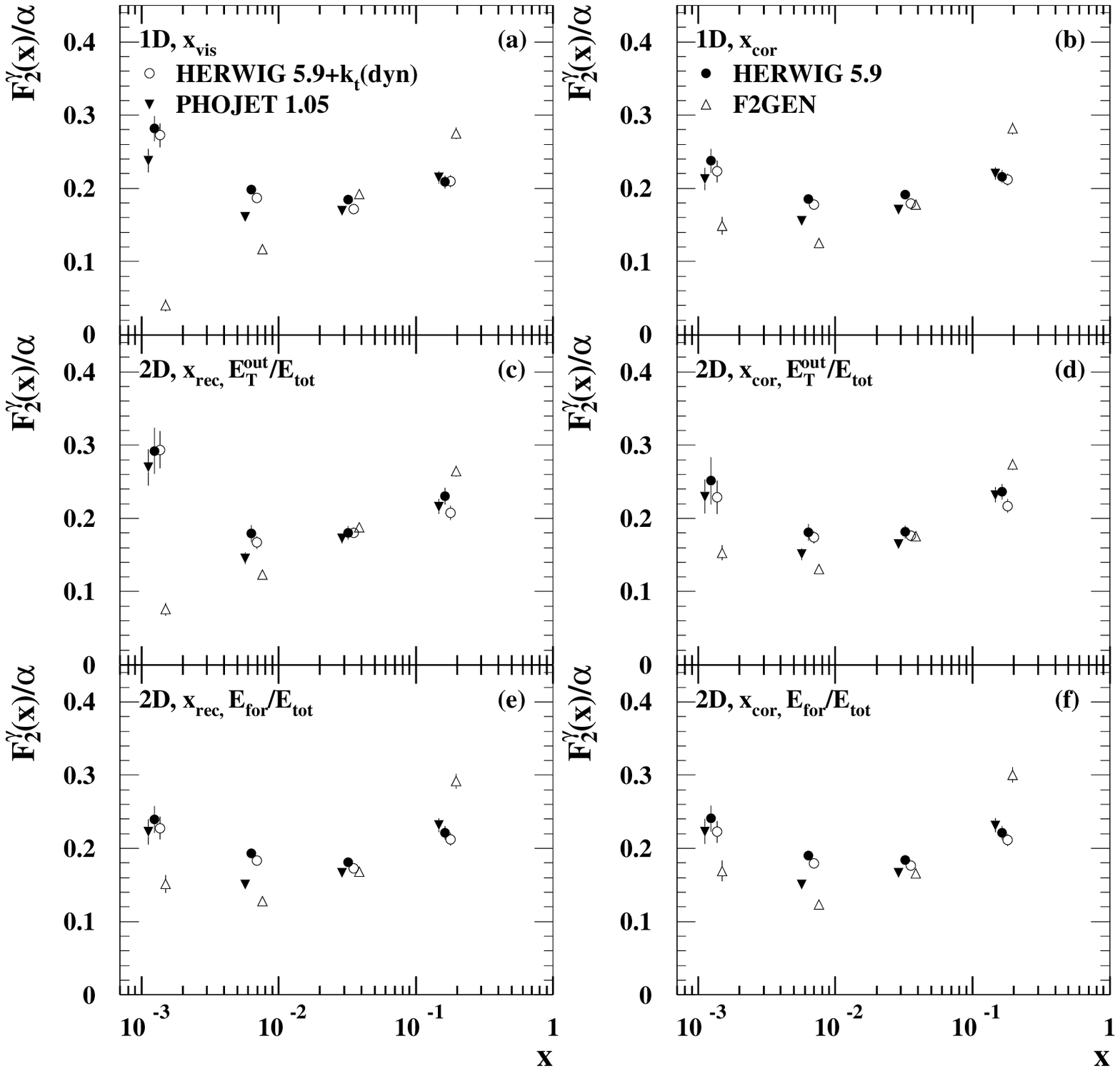,height=28cm}
\vspace{-4.5cm}
\caption{Unfolding of the LEP1 SW low \qsq\ sample with 4 Monte Carlo models and
         different unfolding variables.
         a) \xfd,
         b) \xcor,
         c) \xrec\ and \etdist,
         d) \xcor\ and \etdist,
         e) \xrec\ and \edist,
         f) \xcor\ and \edist.
         The errors are statistical only.
        }\label{91swuf2}
\end{center}
\end{figure}


\begin{figure}[p]
\vspace{-8cm}
\begin{center}
\epsfig{bbllx=50,bblly=0,bburx=595,bbury=842,file=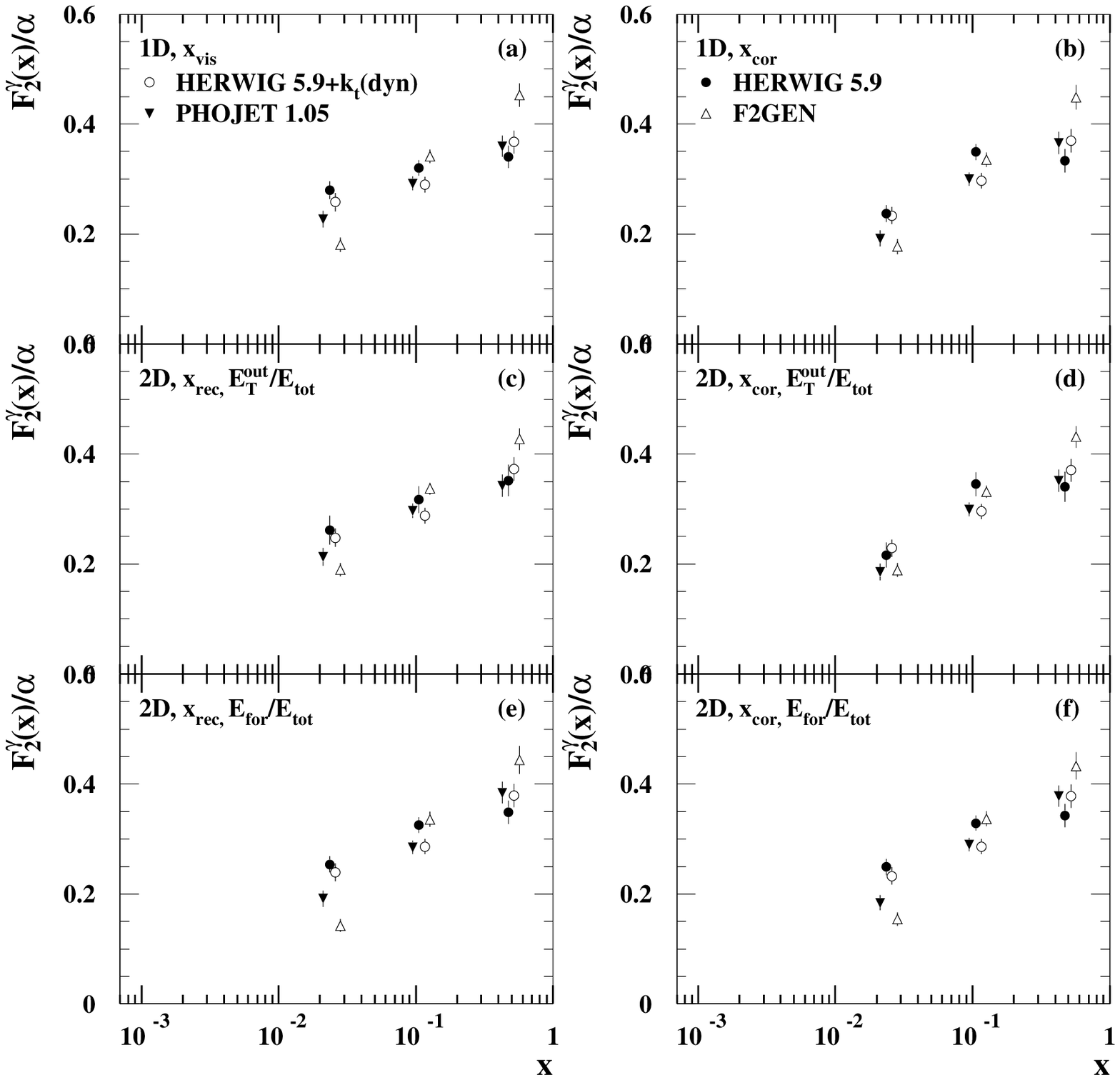,height=28cm}
\vspace{-4.5cm}
\caption{Unfolding of the LEP1 FD low \qsq\ sample with 4 Monte Carlo models and
         different unfolding variables.
         The sub-figures are as described in Figure~\ref{91swuf2}.
         The errors are statistical only.
        }\label{91fduf2}
\end{center}
\end{figure}


\begin{figure}[p]
\vspace{-8cm}
\begin{center}
\epsfig{bbllx=50,bblly=0,bburx=595,bbury=842,file=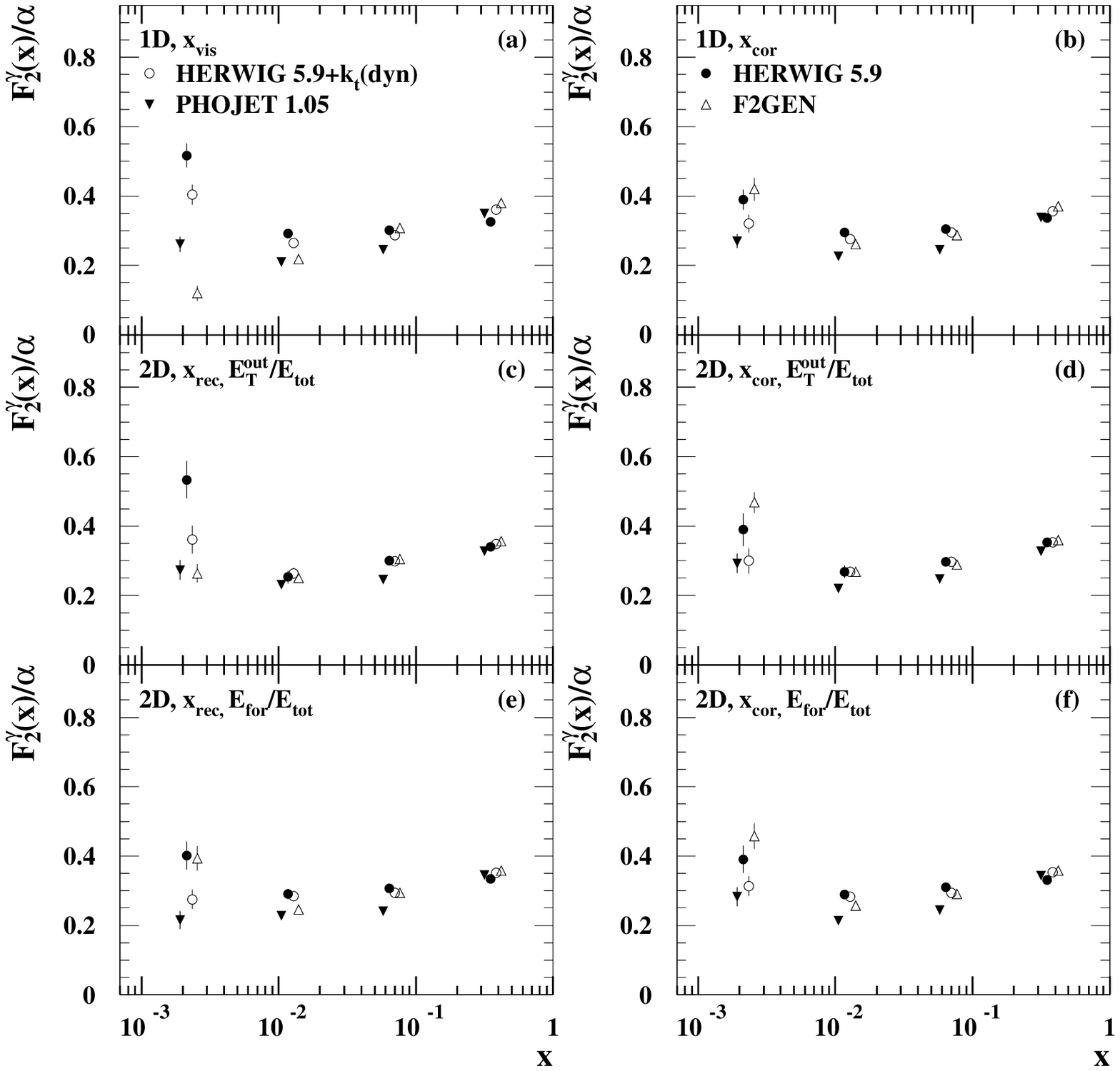,height=28cm}
\vspace{-4.5cm}
\caption{Unfolding of the LEP2 SW low \qsq\ sample with 4 Monte Carlo models and
         different unfolding variables.
         The sub-figures are as described in Figure~\ref{91swuf2}.
         The errors are statistical only.
        }\label{189swuf2}
\end{center}
\end{figure}
\clearpage


\begin{figure}[p]
\vspace{-8cm}
\begin{center}
\epsfig{bbllx=50,bblly=0,bburx=595,bbury=842,file=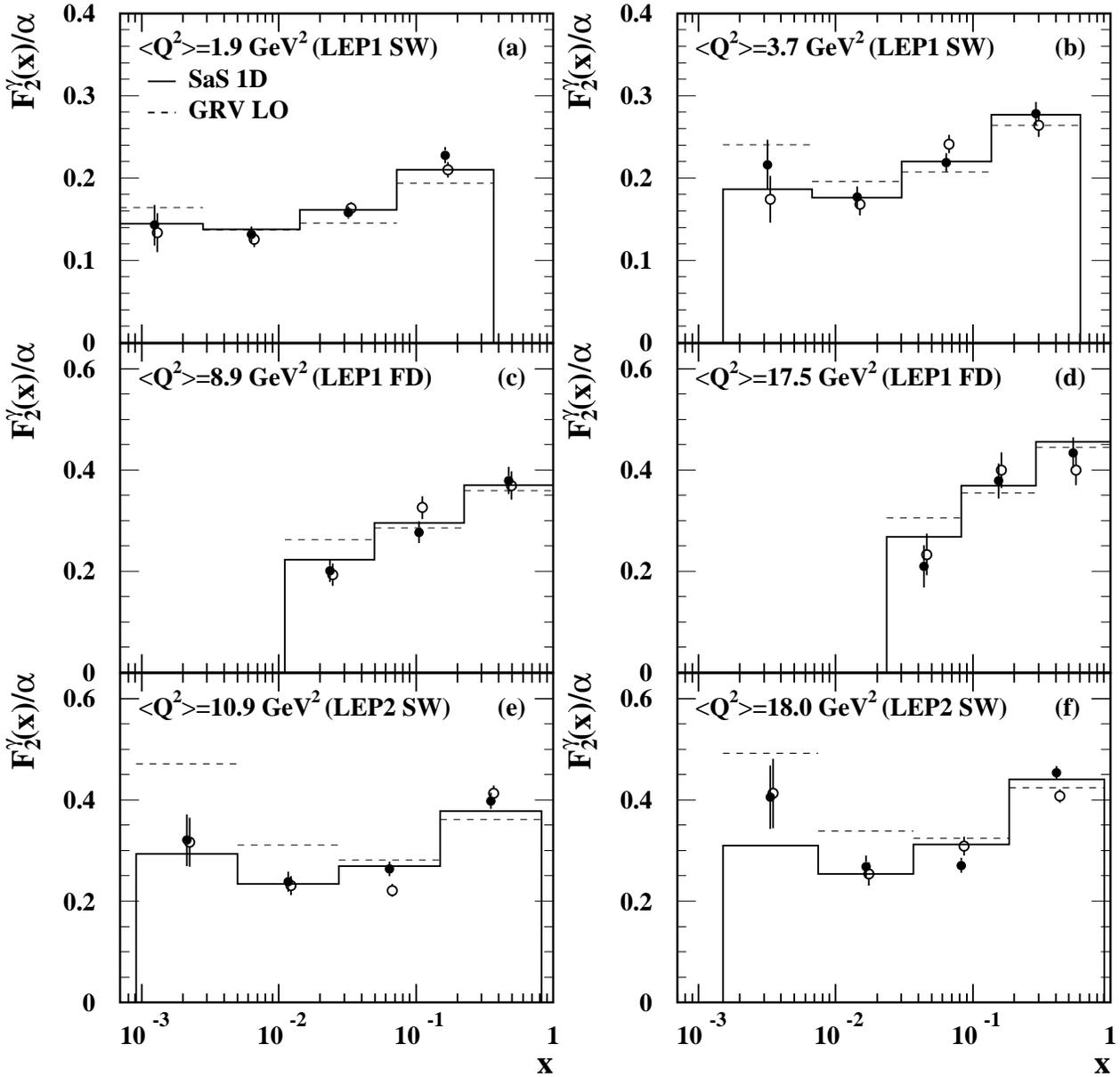,height=28cm}
\vspace{-4.5cm}
\caption{Two independent samples of Monte Carlo events 
         generated using HERWIG 5.9 
         with the SaS1D structure function
         unfolded with HERWIG 5.9 with the GRV LO structure function,
         in six \qsq\ ranges. The solid histogram shows the SaS1D structure
         function at the same average \qsq\ as the sample in each
         plot, weighted by the $x$ distribution in the HERWIG SaS1D
         sample (this is the quantity measured by Equation~\ref{eqn:findft}).
         The dotted lines show the GRV LO structure function weighted
         by the unfolded $x$ distribution. 
         The unfolded SaS1D samples were about the same size as the data
         samples in each \qsq\ region.
         The errors are statistical only.
        }\label{sastest}
\end{center}
\end{figure}
\clearpage


\begin{figure}[p]
\vspace{-8cm}
\begin{center}
\epsfig{bbllx=50,bblly=0,bburx=595,bbury=842,file=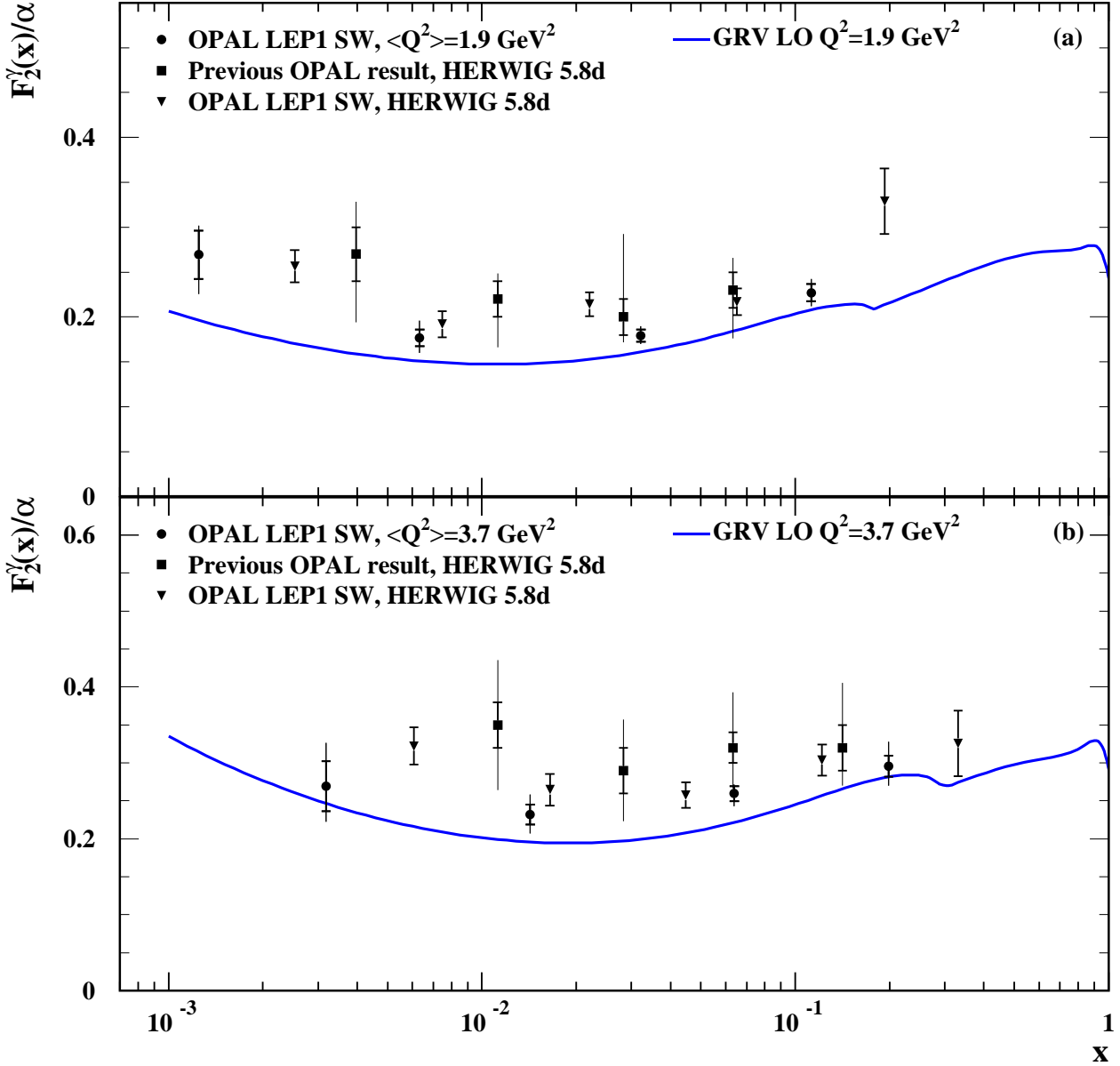,height=28cm}
\vspace{-4.5cm}
\caption{The measurement of \ftn\ using the LEP1 SW sample,
         for \qzm\ values of (a) 1.9 and (b) 3.7~\gevsq.
         Also shown are the previous OPAL results in these \qsq\
         ranges, which were unfolded using HERWIG 5.8d, and
         the result of unfolding the LEP1 SW data 
         using HERWIG 5.8d. 
         For each point, the inner error bars show the
         statistical error and the full error bars show
         the total error, except for the new result with HERWIG 5.8d, 
         for which only statistical errors are shown.
         The positions of the new OPAL points are as given in 
         Table~\ref{tab:resxq}. 
         The other points are shown at the centre of the
         $\log(x)$ bin.
         The curves show the GRV LO structure function.
        }\label{result1opal}
\end{center}
\end{figure}


\begin{figure}[p]
\vspace{-8cm}
\begin{center}
\epsfig{bbllx=50,bblly=0,bburx=595,bbury=842,file=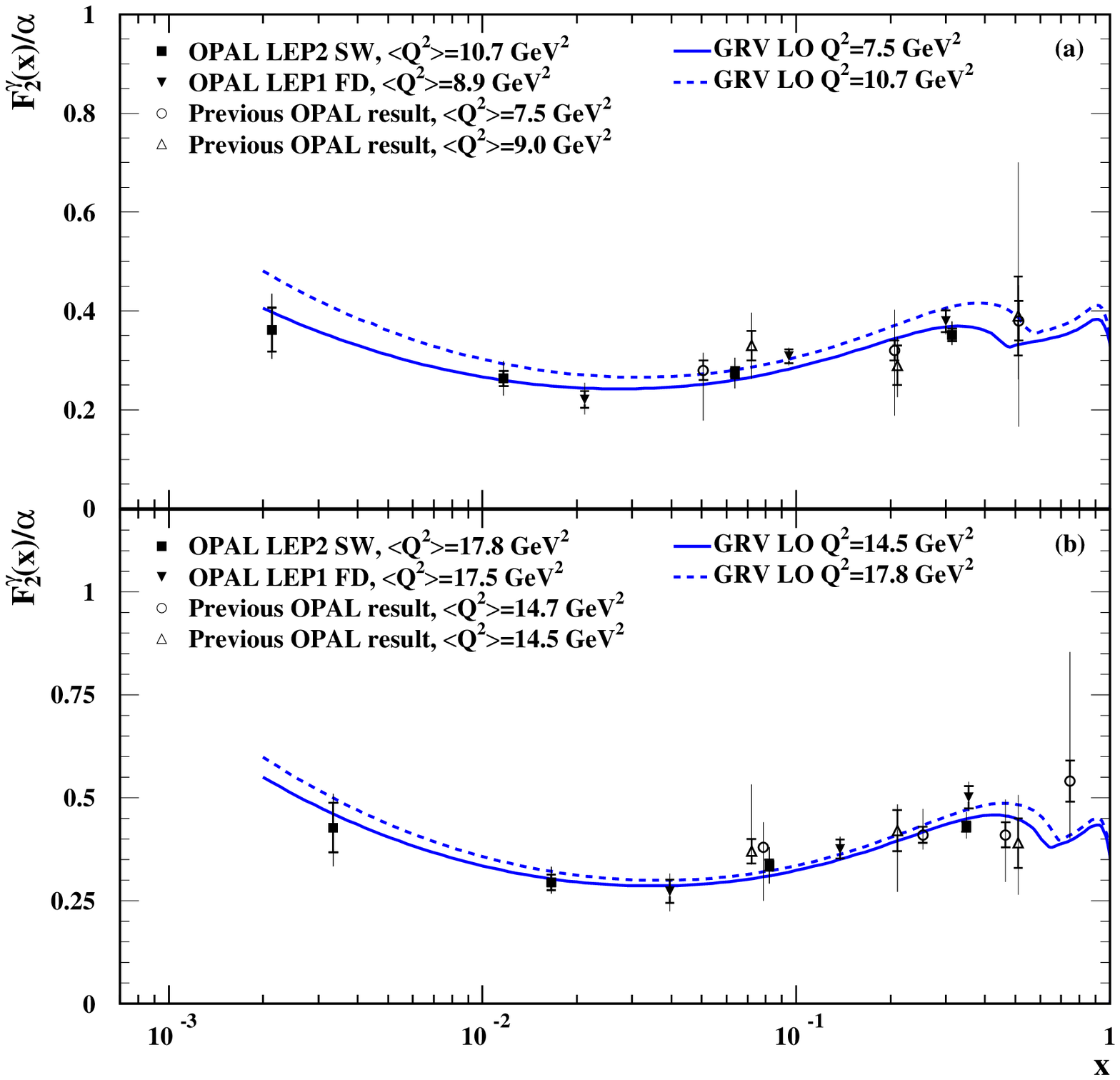,height=28cm}
\vspace{-4.5cm}
\caption{The measurement of \ftn\ using the
         LEP1 FD and LEP2 SW samples
         for \qzm\ values of 
         (a) 8.9 (10.7) and (b) 17.5 (17.8)~\gevsq\
         for LEP1 (LEP2).
         Also shown are the previous OPAL results
         in these \qsq\ ranges, which were unfolded with 
         HERWIG 5.8d (\qzm=7.5 \gevsq\ and \qzm=14.7 \gevsq) 
         and HERWIG 5.9 (\qzm=9.0 \gevsq\ and \qzm=14.5 \gevsq)
         using a linear $x$ scale.
         For each point, the inner error bars show the
         statistical error and the full error bars show
         the total error.
         The positions of the new OPAL points are as given in 
         Table~\ref{tab:resxq}.
         The other points with closed symbols are shown at the centre of the
         $\log(x)$ bin, and those with open symbols are shown at
         the average $x$ value of the bin.
         The curves show the GRV LO structure function.
        }\label{result2opal}
\end{center}
\end{figure}


\begin{figure}[p]
\vspace{-8cm}
\begin{center}
\epsfig{bbllx=50,bblly=0,bburx=595,bbury=842,file=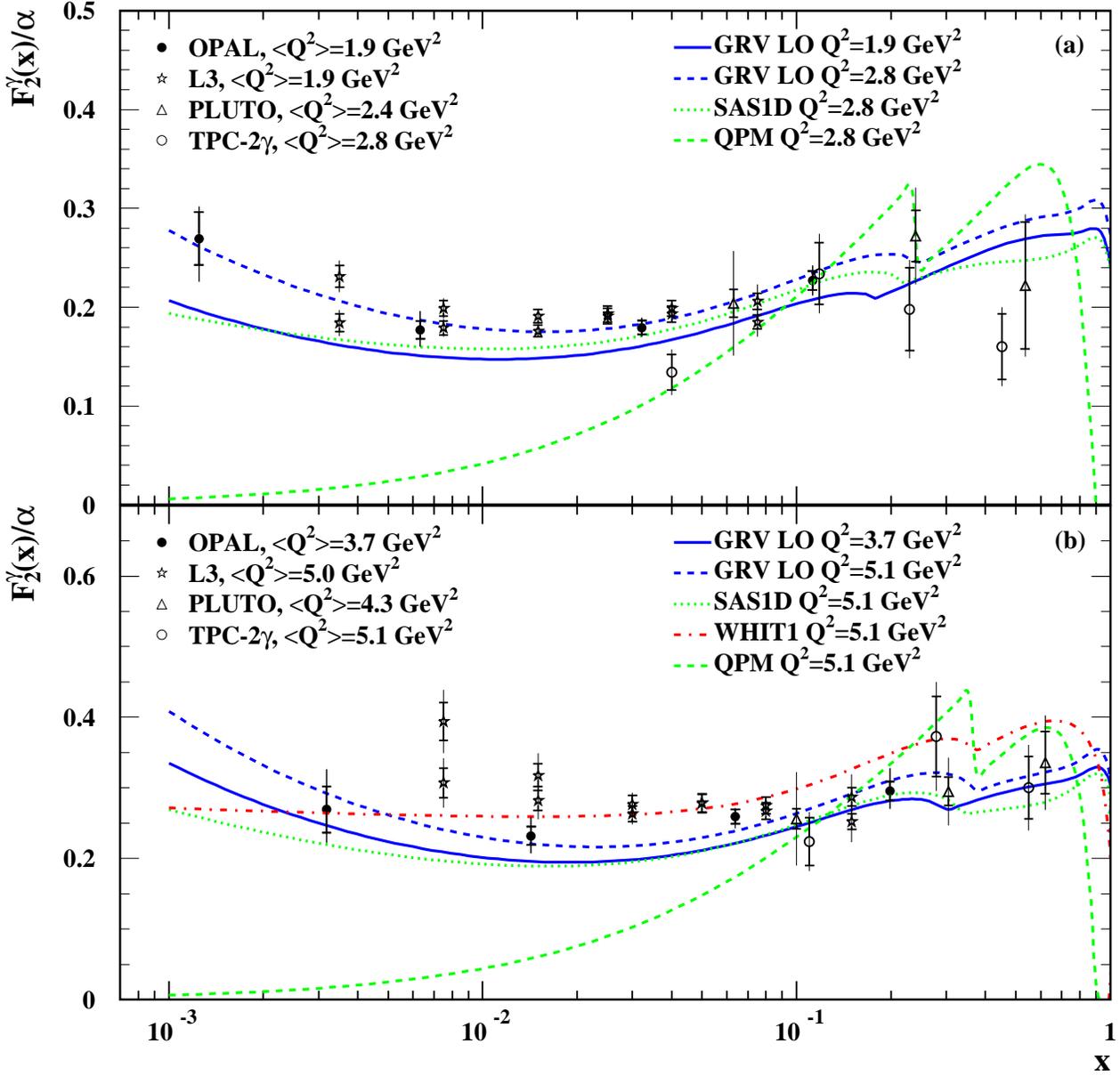,height=28cm}
\vspace{-4.5cm}
\caption{The measurement of \ftn\ using the LEP1 SW sample,
         for \qzm\ values of (a) 1.9 and (b) 3.7~\gevsq.
         Also shown are the results from L3~\cite{ref:L3} 
         PLUTO~\cite{ref:PLUTO}, and TPC/2$\gamma$~\cite{ref:TPC}.
         For L3 the two sets of points were unfolded using different Monte 
         Carlo programs. The lower / upper points correspond to 
         PHOJET 1.05 / TWOGAM.
         For each point, the inner error bars show the
         statistical error and the full error bars show
         the total error.
         The positions of the new OPAL points are as given in 
         Table~\ref{tab:resxq}.
         The other points with closed symbols are shown at the centre of the
         $\log(x)$ bin, and those with open symbols are shown at
         the average $x$ value of the bin.
         The curves show the GRV LO, SaS1D, WHIT1 and QPM structure functions.
        }\label{result1others}
\end{center}
\end{figure}


\begin{figure}[p]
\vspace{-8cm}
\begin{center}
\epsfig{bbllx=50,bblly=0,bburx=595,bbury=842,file=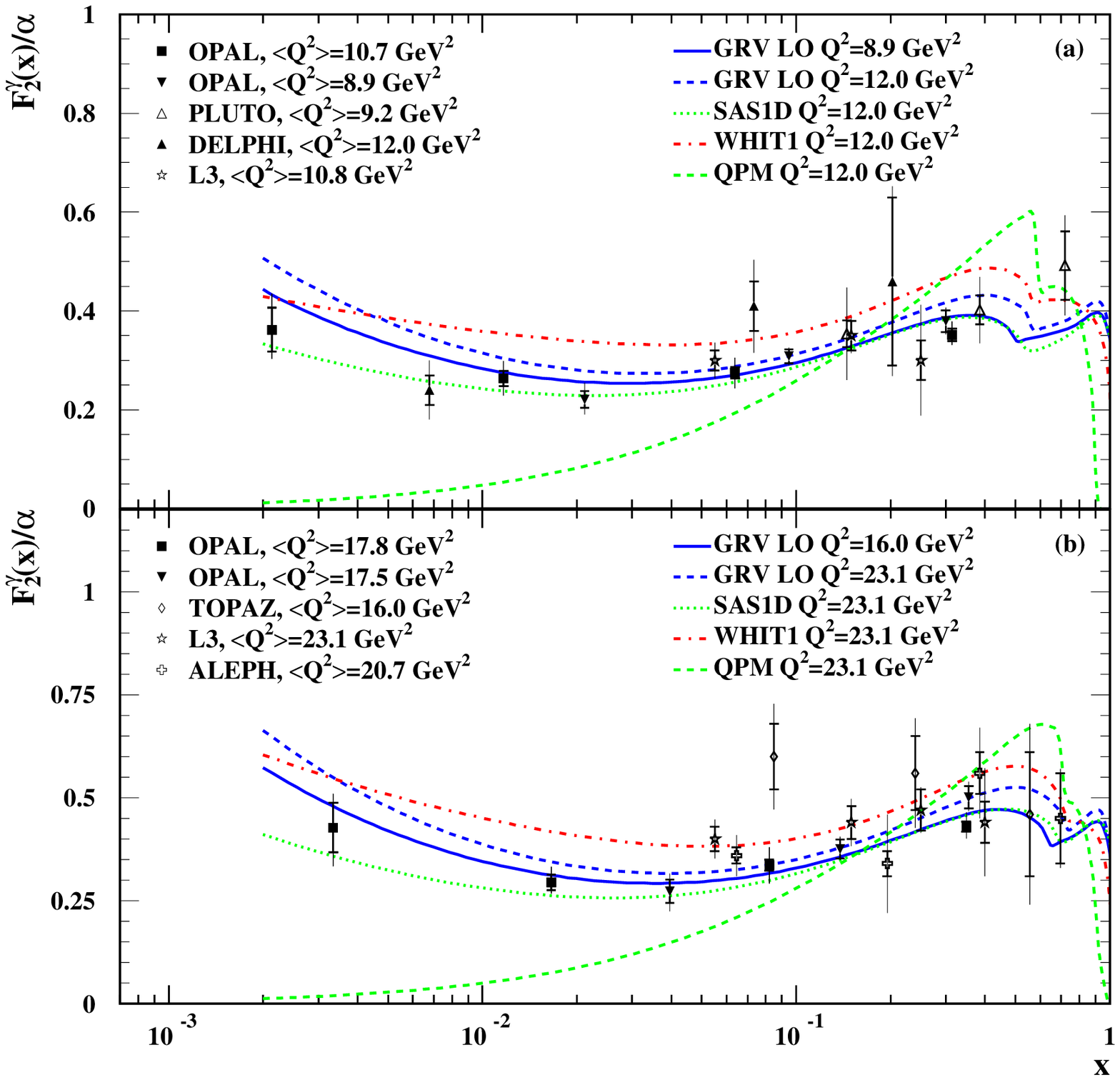,height=28cm}
\vspace{-4.5cm}
\caption{The measurement of \ftn\ using the
         LEP1 FD and LEP2 SW samples
         for \qzm\ values of 
         (a) 8.9 (10.7) and (b) 17.5 (17.8)~\gevsq\
         for LEP1 (LEP2).
         Also shown is a selection of results from other experiments:
         ALEPH~\cite{ref:ALEPH},
         DELPHI~\cite{ref:DELPHI},
         L3~\cite{ref:L3},
         PLUTO~\cite{ref:PLUTO},
         and TOPAZ~\cite{ref:TOPAZ}.
         For each point, the inner error bars show the
         statistical error and the full error bars show
         the total error.
         The positions of the new OPAL points are as given in 
         Table~\ref{tab:resxq}.
         The other points with closed symbols are shown at the centre of the
         $\log(x)$ bin, and those with open symbols are shown at
         the average $x$ value of the bin.
         The curves show the GRV LO, SaS1D, WHIT1 and QPM structure functions.
        }\label{result2others}
\end{center}
\end{figure}


\end{document}